\documentclass[prl,aps,twocolumn]{revtex4}
\usepackage{graphicx}
\usepackage{epsf}
\begin{document}
\title{Quantum charge pumping and electric polarization in Anderson insulators}

\author{Chyh-Hong Chern}
\email{chern@issp.u-tokyo.ac.jp}
\affiliation{Institute for Solid
State Physics, University of Tokyo, Kashiwanoha, 5-1-5, Kashiwa
277-8581, Japan}

\author{Shigeki Onoda and Shuichi Murakami}
\affiliation{CREST, Department of Applied Physics, University of
Tokyo, 7-3-1, Hongo, Tokyo 113-8656, Japan}
\author{Naoto Nagaosa}
\affiliation{CREST, Department of Applied Physics, University of
Tokyo, 7-3-1, Hongo, Tokyo 113-8656, Japan, and Correlated
Electron Research Center, National Institute of Advanced
Industrial Science and Technology, 1-1-1, Higashi, Tsukuba,
Ibaraki 305-8562}
\date{\today}

\begin{abstract}
We investigate adiabatic charge pumping in disordered system in one dimension with open and closed boundary conditions. In contrast to the Thouless charge
pumping, the system has no gap even though all the states are
localized, i.e., strong localization. Charge pumping can be
achieved by making a loop adiabatically in the two-dimensional parameter
space of the Hamiltonian. It is because there are many
$\delta$-function-like fluxes distributed over the parameter space
with random strength, in sharp contrast to the single
$\delta$-function in the pure case. This provides a new and more
efficient way of charge pumping and polarization.
\end{abstract}
\pacs{
77.22.Ej, %Polarization and depolarization (77.22.-d, %Dielectric properties of solids and liquids)
03.65.Vf, %Phases: geometric; dynamic or topological
72.15.Rn, %Localization effects (Anderson or weak localization)
73.63.Nm, %Quantum wires (73.63.-b Electronic transport in nanoscale materials and structures (see also 73.23.-b Electronic transport in mesoscopic systems))
}
\maketitle

\section{Introduction} \label{section:introduction}
The quantum dynamics of the charge in insulators is a rich and
non-trivial issue even in the d.c. limit. Of course, the d.c.
conductivity vanishes, but this does not mean that charge motion is
frozen in the insulators. One example is the ferroelectricity and
electric polarization in insulators. The classical definition of the
electric polarization $\vec{P}=\int d\vec {r} \vec{r} \rho(
\vec{r})$ ( $\rho(\vec{r})$: charge density) fails for the extended
Bloch wave functions since the $\vec{r}$ is unbounded. This
difficulty is avoided by considering instead
the new definition for the {\it difference} of the polarization~\cite{rmp_Resta,Martin,Resta92,Vanderbilt93,OrtizMartin94},
\begin{eqnarray}
\Delta \vec{P} = \int_0^T d \tau  \frac{d\vec{P}}{d \tau},
\label{polarization}
\end{eqnarray}
where the change of the polarization between the initial and final
states are given by the integral of the polarization {\it current}
during the adiabatic change of the parameters $\vec{Q} = (
Q_1,Q_2,\cdots , Q_n)$ such as the atomic displacements. One can usually
choose the initial state with the inversion symmetry without electric
polarization, and (\ref{polarization}) uniquely determines the
polarization of the final state of our interest. Here by using
$\frac{d\vec{P}}{d\tau} = \frac{d\vec{P}}{d Q_i} \frac{d Q_i}{d
\tau}, $ the $\mu$-component of $\Delta {\vec P}$ is expressed as
\begin{equation}
\Delta {P_\mu} = \int_C d {\vec Q} \cdot { {d P_\mu} \over {d
{\vec Q}} }
\end{equation}
with the path $C$ specified by ${\vec Q} = {\vec Q}(\tau)$. What
is found by Resta \cite{Resta92} and King-Smith and Vanderbuilt
\cite{Vanderbilt93} is that ${ {d P_\mu} \over {d {\vec Q}} }$ can
be represented by the Berry phase, which fits to the
first-principle band calculation. Then the question arises, ``Is
there any path ($C$) dependence of the polarization ?". It is
clear that there is no parametrization dependence since we
consider the adiabatic change, but the different path $C_1$ and
$C_2$ in the ${\vec Q}$-space might lead to the different values
of $\Delta {\vec P}$. This is related to the single-valueness and
Chern number of the Bloch wavefunction. Onoda \textit{et al.} addressed
this problem as follows by analyzing the one-dimensional two-band models
characterized by the three dimensional
${\vec Q}$~\cite{OnodaMurakamiNagaosa04,MurakamiOnodaNagaosa07}. There appears a
singular line, i.e., string, in the ${\vec Q}$-space corresponding
to the trajectory of the monopole (band crossing point) as the
momentum $k$ moves, which acts as the ``current circuit" to
produce the ``magnetic field" ${ {d P} \over {d {\vec Q}} }$ via
the Biot-Savart law. Away from the string, the system is always
gapped, and insulating. When the adiabatic change of the parameter
${\vec Q}$ makes a loop enclosing the string, the charge is pumped
during this process, which is quantized to be an integer multiple
of $e$ since the strength of the ``current" is quantized. This is
a realization of the quantum charge pumping first proposed by
Thouless \cite{Thouless83}. This analogy to the magnetostatics says that the
polarization is path independent as long as the loop $C = C_1 +
(-C_2)$ does not enclose the string. This is usually the case
because the change of the parameter ${\vec Q}$  is rather small
and we need the huge variation of ${\vec Q}$ to enclose the
string, i.e., gapless states. In other words, the ferroelectricity
can be regarded as ``a fraction of the quantum charge pumping".

This charge pumping can be regarded as the rigid shift of the
wavefunction due to the change in the external
parameters such as the atomic positions. It is natural when the
wavefunction consists of the Bloch states extended over the whole sample,
and then the quantum interference pattern is modified by the external
parameters. One can estimate roughly how much the charge is pumped
as below. Let the dimension of the parameters be the energy. Then
the distance between the physically realized set of parameters and
that of gapless states, i.e., string, is the energy gap $E_G$.
When $\Delta$ be the change of the the parameters in unit of
energy, the angle subtended by this {\it segment} in parameter
space is roughly estimated as $2 \pi \Delta/E_G$. Since the $2
\pi$ winding corresponds to unit charge $e$ shift by one lattice
constant $a$, i.e., $P_0 = ea$, the polarization is roughly given
by $P = ea \Delta /E_G$~\cite{OnodaMurakamiNagaosa04}. Therefore
one can enhance the dielectric response by reducing $E_G$, or
enlarging $\Delta$. One possible method to reduce $E_G$ is by
introducing disorder, by which even the gapless insulator can be
realized. However, the electron wavefunctions are no longer the
extended Bloch states in the presence of the disorder, and one
needs to worry about the localization, i.e, Anderson localization.
When all the states are strongly localized, one can not
transmit the phase information though the sample, and hence can not expect
the charge pumping either. On the other hand, as can be seen from the above explanation, the charge pumping is closely related to the topological nature
of the wavefunctions in the parameter space, which is robust against the disorder to some degree. For example, in the two dimensional electron systems under strong magnetic field, there occur discrete extended states {\it protected by the topology}. Namely, the Chern number is carried by the extended states only, which is not destroyed by the weak disorder. Therefore similar problem arises even for the one-dimensional systems which we study below.
Niu and Thouless \cite{NiuThouless84}
was the first to study the stability of the charge pumping
against the weak disorder  by a topological argument.
Even though all the states are localized, the charge pumping
was shown to be unchanged as long as the gap remains finite.
However, questions still remain such as what is the physical mechanism
of the charge pumping through the localized states and
what happens when the gap collapses.

In this paper, we study the effect of the disorder on the charge
pumping and dielectric response in a one-dimensional model for
insulators. Combining the numerical simulations and analytic
considerations, we reveal the physical picture of the charge
pumping by the localized states both in the case of open boundary
condition and periodic/twisted boundary condition. The former one
is more relevant to the experimental situation such as the FeRAM
(ferroelectric random access memory)
where the leads are attached to the thin film of insulators, while
the latter is more appropriate to see the role of topology. We
have published a Letter focusing on the open boundary condition
\cite{onoda2006prl}. This paper provides the full details of the
formulation, calculations, and additional results on the open
system, as well as new results on the periodic/twisted boundary
condition.

  The plan of this paper follows. In section II, a model for disordered
insulator showing the charge pumping is introduced. Its analysis
with the open boundary condition is given in section III,
including the detailed description of the resonant tunnelling
mechanism.  In section IV given the analysis of the model in the
three dimensional parameter space including the angle $\alpha$ for
the twisted boundary condition, and the role of magnetic monopoles
in this space. Section V is devoted to the conclusions.

\section{A model for disordered insulator}

The minimal model for ferroelectrics is given by the following ionic dimer model;
\begin{eqnarray}
  H_{\mathrm{pure}} &=& -\frac{1}{2}\sum_{i=1}^L\left(t_{\mathrm{n.n.}}-(-)^iQ_2\right)
  (c^\dagger_{i+1}c_i+h.c.)
  \nonumber\\
  &&\makebox[1cm]{}
  +\sum_{i=1}^L (-)^{i}Q_1 c^\dagger_ic_i.
  \label{eq:H_pure}
\end{eqnarray}
Here, $c_i$ and $c_i^\dagger$ are the annihilation and creation operators
of the electron at the site $i=1,\cdots,L$ with $L$ being the number
of sites in the system. For open system attached to leads at the both ends,
$c_{L+1}$ and $c^\dagger_{L+1}$ represents those operators in one of the leads,
while for closed system, they are understood as $c_1$ and $c^\dagger_1$.
$t_{\mathrm{n.n.}}$ is the transfer integral.
$Q_1$ and $Q_2$ represent the alternations of the local ionic level
and the bond dimerization, respectively.
The spin degree of freedom is omitted for simplicity.
Then, of our interest is the half-filling case relevant to the ferroelectrics.
Although this model might look special, it represents the two essential
feature of the ferroelectrics, i.e., (i) the two species of the ions
characterized by the level alternation $Q_1$ and (ii) relative shift of the atomic positions
described by the dimerization $Q_2$.  It corresponds to the
ferroelectricity in e.g. BaTiO$_3$, where Ti and O
are dimerized to produce the polarization \cite{Ishihara93}.
It can be also applied to the quasi-one dimensional ferroelectric
materials such as organic charge transfer compounds
TTF-CA~\cite{ni} and (TMTTF)$_2$X~\cite{tmttf}.

The Hamiltonian $H_{\mathrm{pure}}$ under the periodic boundary
condition yields two bands\cite{OnodaMurakamiNagaosa04}
\begin{equation}
  \varepsilon_{\pm}(k) = \pm \sqrt{
    t_{\mathrm{n.n.}}^2\cos^2 k + Q_1^2 + Q_2^2 \sin^2 k}.
  \label{eq:bands}
\end{equation}

\begin{figure}[ht]
  \begin{center}\leavevmode
    \includegraphics[width=8.4cm]{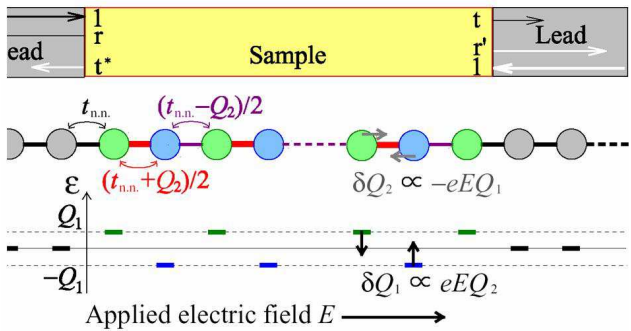}
  \end{center}
 \caption{\label{fig:model}(Color online) Ionic dimer system sandwitched by
the leads.}
\end{figure}

Experimentally, the parameters $\vec{Q}\equiv(Q_1,Q_2)$ can be controlled by applying the electric field $E$ along the polarization direction and the pressure $p$ as follows. Electrons at high and low deinsity sites shift in relatively opposite directions, as shown in open arrows of Fig.~\ref{fig:model}~(b), changing the dimerization $Q_2$ by $\delta Q_2\propto eEQ_1$. Simultaneously, within each dimer, a level difference $Q_1$ changes by $\delta Q_1\propto-eEQ_2$ as illustrated by black arrows in Fig.~\ref{fig:model}~(b).
Therefore, the electric field $E$ mainly controls the angle
\begin{equation}
  \theta=\arctan(Q_2/Q_1).
  \label{eq:theta}
\end{equation}
Applying the pressure, one can increase the hybridization, and then reduces the ratio $Q/t_{\mathrm{n.n.}}$ with
\begin{equation}
  Q=\sqrt{Q_1^2 + Q_2^2}.
  \label{eq:Q}
\end{equation}

\begin{figure}[htb]
\begin{center}
  \includegraphics[width=8cm]{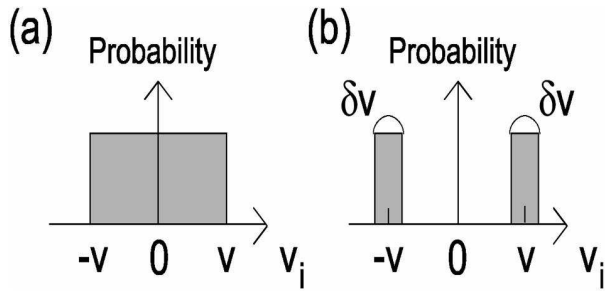}
\end{center}
\caption{(a) Uniform distribution of the on-site random potential. (b) Alloy model for the on-site random potential, which mimics effects of substitution in alloys.}
\label{fig:random}
\end{figure}

To discuss the quantum relaxor behavior, we introduce the on-site random potential $v_i$ to the Hamiltonian given by (\ref{eq:H_pure});
\begin{equation}
  H = H_{\mathrm{pure}}+\sum_{i=1}^L v_i c^\dagger_i c_i.
  \label{eq:H}
\end{equation}
The type of the random distribution takes either a uniform distribution or an alloy model, which are shown in Fig.~\ref{fig:random}. In the following, we study effects of on-site disorder on dielectric properties in both open and closed systems, particularly focusing on the topological aspects.

\section{Open boundary condition}

In this section, dielectric property of disordered insulators
connected to the leads is studied theoretically. This corresponds
to the realistic situation in FeRAM devices, where the
polarization is measured by the integrated current flowing in the
leads. Especially in nano-scale samples of the FeRAM, the quantum
nature of the polarization is expected to play essential roles via
novel quantum interference as discussed in the
introduction~\cite{Thouless83,OnodaMurakamiNagaosa04}. Namely, a
theory for the nano-scale capacitance is called for where the
quantum coherence is attained all through the sample.
>From the viewpoint of the applications, it is required to achieve (i) a magnitude of the polarization larger than 10 $\mu$C/cm$^2$, (ii) a smaller leak current than 0.1-1 $\mu$A/cm$^2$, and (iii) a dielectric constant larger than 300~\cite{Scott}.
By introducing disorder, one can also suppress the
dissipation by leak current due to the strong localization effect~\cite{LeeRMP,NiuThouless84} in addition to the reduction of the
gap and enhanced dielectric response. In fact, a similar disorder-driven enhanced response has been found in relaxor ferroelectrics~\cite{Cross87,science_relaxor,Horiuchi00} and pinned charge density waves~\cite{GrunerRMP}, although its mechanism is due to a classical mesoscopic cluster formation analogous to spin glass systems.

In the disordered case with all the states being localized, the
charge transfer occurs  through the resonant tunnelling.
%In the single-channel problem,
The quantum nature is topologically protected. Crossover from the quantized charge pumping in bulk insulator~\cite{Thouless83} to that of quantum dot~\cite{Marcus99} is clarified in the control parameter plane as a function of sample size and disorder strength~\cite{LevinsonEntinWohlmanWolfre01,EntinWohlmanAharony02,KashcheyevsAharonyEntinWohlman04}. Then, we theoretically propose the enhancement of the dielectric response of insulators by disorder in nanoscopic/mesoscopic multichannel systems, with the required time for the dissipationless adiabatic charge displacement being estimated. Application of this phenomenon to memory devices, like ferroelectric random access memory (FeRAM)~\cite{Scott}, is also discussed.

\subsection{Single-channel problem for uniformly distributed random potential}
\label{subsec:single}
We consider an insulating electronic system sandwiched by leads
(electrodes) as shown in Fig.~2(a). For simplicity, we take a
one-dimensional (single-channel) model, but the extension to
higher dimensional (multi-channel) cases is straightforward as
discussed later.

We take the total Hamiltonian
\begin{equation}
  H_{\text{tot}} = H + H_{\mathrm{lead}}
  \label{eq:H_tot}
\end{equation}
with $H$ and $H_{\mathrm{lead}}$ being given by (\ref{eq:H}) and
\begin{eqnarray}
  H_{\mathrm{lead}} &=& -\frac{t_{\mathrm{n.n.}}}{2}\left(\sum_{i=0}^{-\infty}+\sum_{i=L}^{\infty}\right)
  \left(c^\dagger_{i+1}c_i+h.c.\right),
  \label{eq:H_lead}
\end{eqnarray}
respectively. This model is schematically shown in Fig.~\ref{fig:model}~(b).
The Green's function $G_{i,i'}(\varepsilon)$ for the above model is readily obtained from the recursion formula in the form of the continuued fraction~\cite{LandauerButtiker}. Here, we can concentrate on the case where the chemical potential is located at the zero energy.

We adopt the Landauer-B{\"u}ttiker formalism~\cite{LandauerButtiker,Aharony},
where the sample is regarded as a {\it scatterer} characterized by
the scattering $S$-matrix
\begin{equation}
  S = \left(\begin{array}{cc}
    r & t \\
    t^* & r'
  \end{array}\right).
  \label{eq:S}
\end{equation}
Here, the reflection coefficients $r$ and $r'$, and the transmission coefficient $t$ (see Fig.~\ref{fig:model}) can be calculated from the Green's functions~\cite{LandauerButtiker}.
The transmittance through the sample and the reflectance at the
both ends are expressed as
\begin{eqnarray}
  T &=& |t|^2,
  \label{eq:T}\\
  R &=& |r|^2 = |r'|^2=1-T,
  \label{eq:R}
\end{eqnarray}
respectively. Then, we employ the Brouwer's
formula~\cite{Brouwer98,Buttiker94}, which has been successfully
applied to the charge pumping in metallic quantum dot
systems~\cite{Marcus99}: the charge $\Delta q$ pumped from the
left to the right during an adiabatic change of parameters
$\vec{Q}$ along a path $C$ is given by ~\cite{Brouwer98}
\begin{equation}
 \Delta q = e\int_C\!\frac{d \vec{Q}}{2\pi} \cdot {\rm Im}
 \left(r^*\vec{\nabla}_Q r + t\vec{\nabla}_Q t^*\right).
 \label{eq:q}
\end{equation}
Therefore, even in the perfectly reflecting case $t=0,|r|=1$,
charge can be pumped by controlling the phase $\varphi$ of $r=e^{i\varphi}$~\cite{Avron2004}.

\begin{figure}[tbp!]
  \begin{center}\leavevmode
    \includegraphics[width=8.0cm]{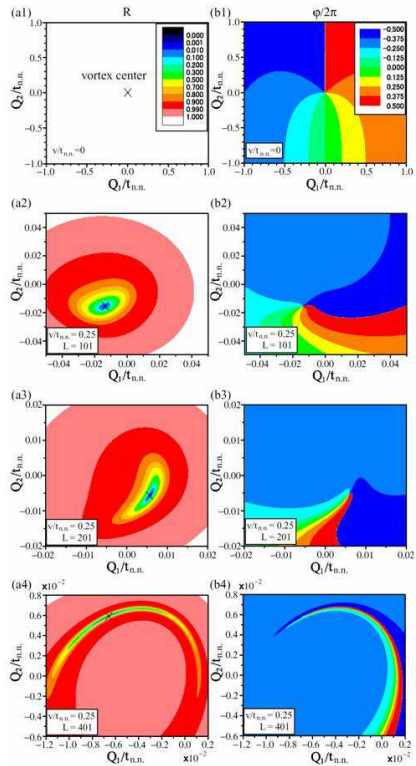}
  \end{center}
 \caption{\label{fig:r-phase} (Color) Reflectance $R=|r|^2$, and the phase
$\varphi={\rm arg}\ r$. (a1), (b1): clean bulk system. (a2), (b2):
disordered system of finite size $L=401$ with the random potential
of the strength $v/t_{n.n}=0.5$. The color code given in (a1)/(b1)
also applies to (a2,3)/(b2,3). In the white region in (a1,2,3),
$R$ is the unity within the accuracy of $10^{-4}$.}
\end{figure}

Let us start with the pure case ($v_i=0$). Without the random
potential $v_i$, the bulk system has an energy gap
\begin{equation}
  E_{G0} = 2 \sqrt{ Q_1^2 + {\rm Min}\{t_{\mathrm{n.n.}}^2,Q_2^2\}}.
  \label{eq:E_G0}
\end{equation}
This gap closes at $\vec{Q}={\vec 0}$, where the metallic
conduction occurs. This yields a vortex center of the reflective
coefficients $r$ and $r'$. By means of the continuued-fraction
expansion of the Green's function, an analytic form of $r$ in the
thermodynamic limit $L \to \infty$ is obtained as
\begin{equation}
  r = -\frac{Q_1^2+t_{\mathrm{n.n.}}Q_2+Q\sqrt{t_{\mathrm{n.n.}}^2+Q_1^2}+2it_{\mathrm{n.n.}}Q_1}
  {Q_1^2+t_{\mathrm{n.n.}}Q_2+Q\sqrt{t_{\mathrm{n.n.}}^2+Q_1^2}-2it_{\mathrm{n.n.}}Q_1}
  \label{eq:r_pure}
\end{equation}
where $Q\equiv\sqrt{Q_1^2+Q_2^2}$. The panels (a1) and (b1) of
Fig.~\ref{fig:r-phase} show the reflectivity $R=|r|^2$ for
$L=10001$ and the phase $\varphi$ of $r$ for $L\to\infty$ without
randomness. Here the phase winds by $2 \pi$ around
$\vec{Q}=(0,0)$, forming a ``vortex'' at which $r=0$ and $|t|=1$.
In the thermodynamic limit $L\to\infty$, $R=1$ and $T=0$ hold
except at the vortex center. For finite-size systems, the
transmittance $T$ behaves as $e^{- L/\xi_{0}}$ with $\xi_0 =
t_{\mathrm{n.n.}}/E_{G0}$. Therefore the size of the region in the
$\vec{Q}$ space with a large $T$ is of the order of
$t_{\mathrm{n.n.}}/L$ for a large $L$. Namely the vortex
corresponds to the gapless case, where the extended state at the
Fermi energy carries the charge and causes the perfect
transmittance. Remarkably, this perfect-transmission point
$\vec{Q}=\vec{0}$ governs topological properties of the system in the
whole $\vec{Q}$ plane in a non-local way. When we adiabatically
change the parameter $\vec{Q}$ along a cycle around $\vec{Q}={\vec
0}$ at which the transmittance $T$ vanishes, the charge $e$ is
pumped and the polarization changes by $\pm 2ea$ ($a$: lattice
constant) according to the Brouwer's formula ~(\ref{eq:q}). The
vortex is almost isotropic at least in the vicinity of its core.
Then, a pumped charge due to a small change ${\vec \Delta}$ of
$\vec{Q}$ is expressed as $q\sim(\phi/2\pi)
e\sim(|\vec{\Delta}|/E_G)e$, where $\phi$ represents a change in
the polar angle of the vector $\vec{Q}$. Therefore one can enhance
the pumped/displaced charge by enlarging the angle $\phi$
subtended by ${\vec \Delta}$ around the gapless point. This
quantized charge pumping~\cite{Thouless83} through the adiabatic
cyclic change of $\vec{Q}$  is consistent with the results
previously found in the periodic
system~\cite{OnodaMurakamiNagaosa04} by using the Berry-curvature
formulation of the electric polarization~\cite{rmp_Resta}.

Now we turn on the disorder in order to reduce the gap $E_G$.
In the ${\vec Q}$ plane, the region of $|{\vec Q}| \lesssim v$
becomes gapless, while the Anderson localization seriously affects
the transport properties~\cite{Anderson58,LeeRMP}.
In one dimension, the effect is pronounced and all the states are
localized~\cite{Wegner76,weaklocalization79,MacKinnonKramer81}.
\begin{figure}[htb]
  \begin{center}
(a)
    \includegraphics[width=7.0cm]{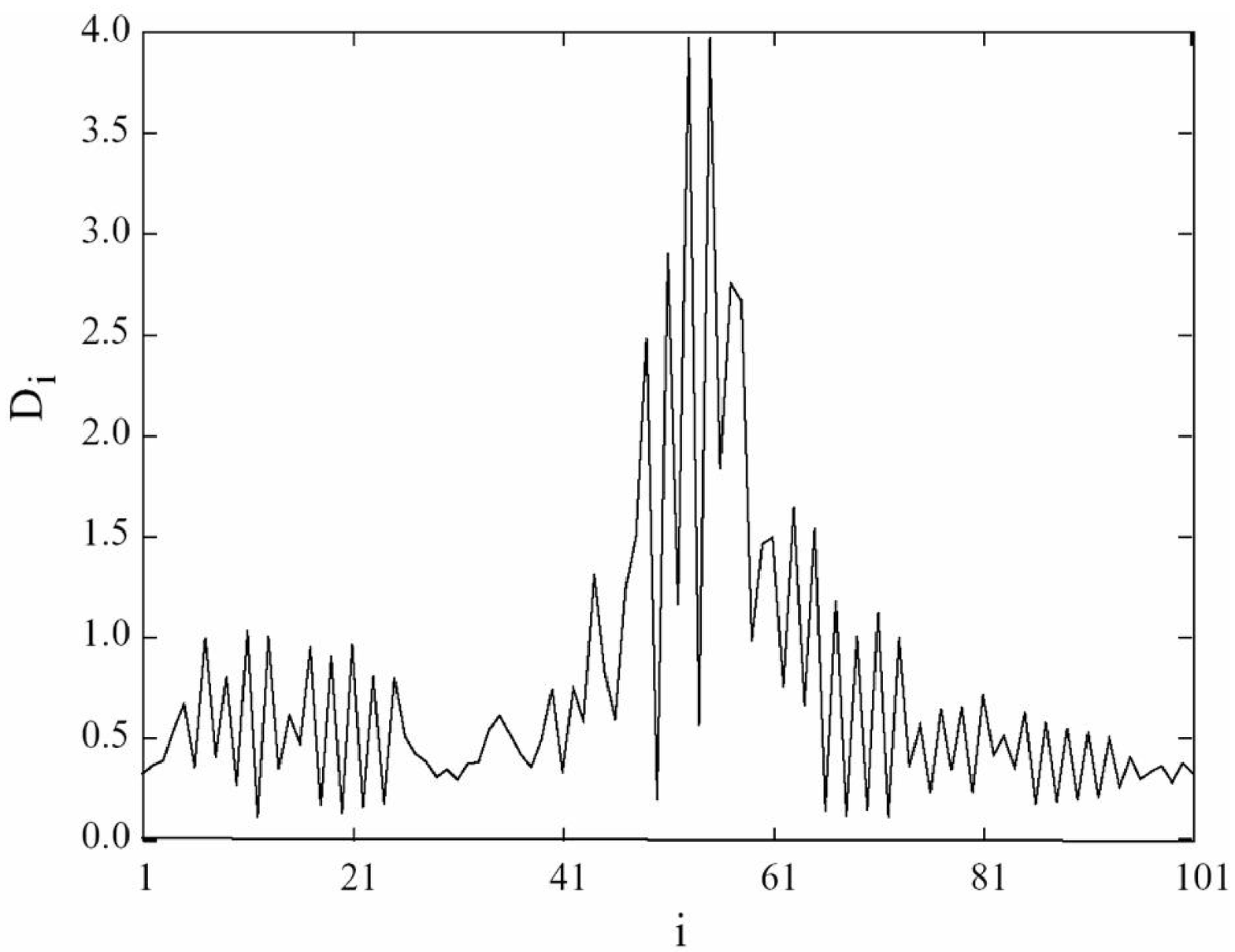}

(b)
    \includegraphics[width=7.0cm]{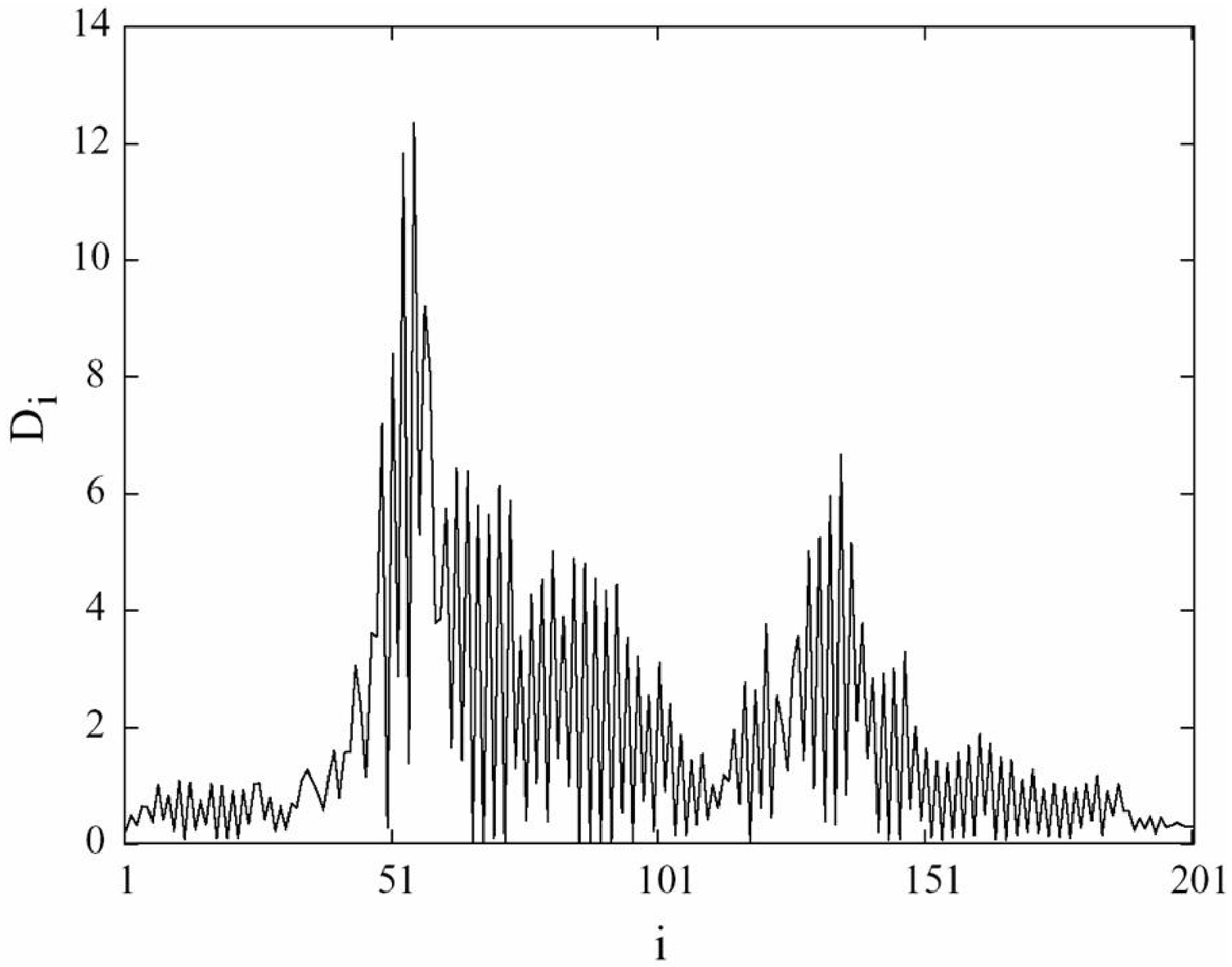}

(c)
    \includegraphics[width=7.0cm]{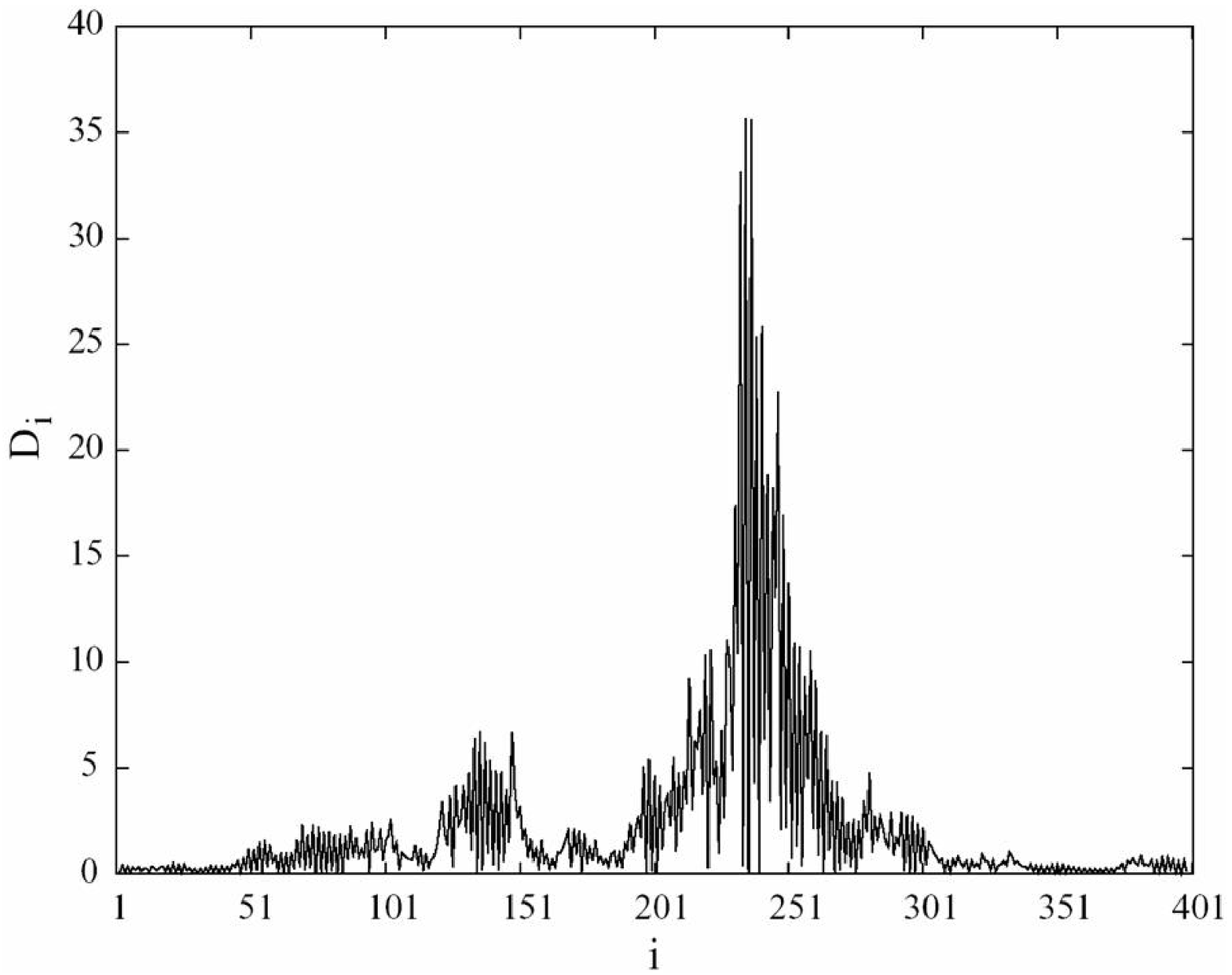}
  \end{center}
  \caption{Local probability $D_i$ of the state at the vortex
for (a) $L=101$, (b) $L=201$, and (c) $L=401$ in the presence of uniformly distributed random potential $v_i\in[-v,v]$ with $v=0.25t_{n.n.}$.}
  \label{fig:ldos}
\end{figure}
On the other hand, the total vorticity $N_v$ is an integer
topological number, and is robust under a continuous change of
parameters including the disorder strength. Therefore, even with
disorder, $N_v$ remains unity. Namely, there should exist at least
one vortex ($r=0$) in the $\vec{Q}$ plane; thus perfect
transmittance $|t|=1$ occurs even though all the states are
strongly localized. This is a remarkable consequence from the
quantum and topological property of the system, and is in sharp
contrast to usual classical tunnelling through disordered
insulators.

In Figs.~\ref{fig:r-phase} (a2) and (b2), we show our numerical results
of $R$ and $\varphi$ for $L=101$
with a uniform random distribution $v_i\in[-v,v]$ for
the disorder strength $v=0.25t_{\mathrm{n.n.}}$. The vortex center of the
perfect transmittance $T=1$ and $R=0$ shifts in the $\vec{Q}$ space.
Besides, an anisotropy develops in the shape of the region of relatively high transmittance $T$. We further calculate $R$ and $\varphi$ for larger systems, which are shown in (a3) and (b3) for $L=201$ and in (a4) and (b4) for $L=401$.
There, it is evident that the vortex core, which is almost isotropic without the disorder, rapidly evolves into the highly anisotropic one. This is associated with an increase of the ratio $L/\xi$ by increasing the system size. Figure~\ref{fig:ldos} shows the local density of states
$D_i$ of site $i$($=1,\cdots,L$), which has been calculated when $\vec{Q}$ is located at the vortex center $\vec{Q}_c$ corresponding to each case of $L=101$, 201 and 401 for $v/t=0.25$. When $L$ is equal to $101$ or smaller, the state at this energy is extended over the sample. However, with increasing $L$, the state is almost localized in the middle of the sample. The spatial extension of the wave function, i.e., the localization length $\xi$, can be explicitly evaluated from the second moment of
$D_i$ as the inverse participation ratio;
\begin{equation}
  \xi_{\mathrm{IPR}} = (\sum_i D_i)^2/\sum_i D_i^2.
  \label{eq:xi_IPR}
\end{equation}
We obtain $\xi_{IPR}=51.4$, 80.8, 76.4 and 84.7 for $L=101$, 201, 401 and 501, respectively around the vortex center, indicating that $\xi_{IPR}$ almost saturates about 80 sites for $v/t=0.25$.

\begin{figure}[htb]
  \begin{center}
    \includegraphics[width=7.5cm]{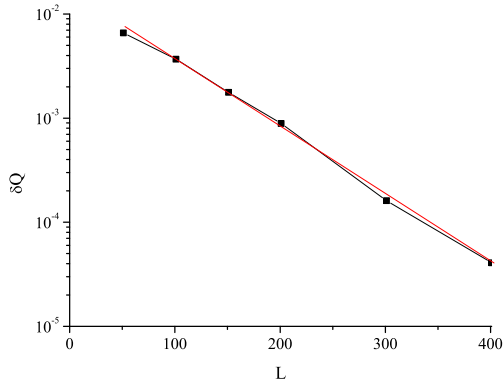}
    \caption{\label{fig:width}The minimum width $\delta Q$ for $R$ determined at $R=0.3$ as a function of $L$.}
  \end{center}
\end{figure}

In this anisotropic ``wing", $\varphi$ changes rapidly. The width
$Q_W$ of the anisotropic wing decays exponentially as
$\exp(-L/\xi)$ with increasing $L$, as shown in
Fig.~\ref{fig:width}.

In the rest of this subsection, we comment on the relation of the
charge displacement discussed above to the charge pumping through
a {\it metallic dot}~\cite{Marcus99} weakly connected with two
leads. In the metallic dot systems, a charge is pumped from one
lead to the other by changing the heights of potential barriers
between the dot and the neighboring leads~\cite{Brouwer98}. Since
the system is metallic, the pumped charge is not quantized in this
case. Here, the main source of the pumping is a finite
transmittance but not the phase of reflection coefficient. With
increasing the system size, it crosses over to the opposite regime
with vanishing transmittance studied in this paper. In the present
theory, the charge pumping comes from the vortex of the reflection
coefficient. Therefore, it is topologically protected against
disorder. This topological constraint guarantees the applicability
of the resonance tunnelling \cite{Azbel1983}, only near the
vortex.

\subsection{Resonance tunnelling}
\label{section:resonance-open} In order to understand transport
properties of such one-dimensional disordered systems, it is
sometimes useful to consider an effective model with the potential
having high double peaks. In this model, well-defined localized
eigenstates exist between the two potential peaks, and any
transport between two ends of the system occurs via the localized
states through tunnelling. While such tunnelling has exponentially
small probability, it occurs when the Fermi energy of the leads is
equal to one of the eigenenergies of the localized states, namely
when the resonance takes place. This picture of transport in
disordered systems is called ``resonance tunnelling''
\cite{Azbel1983}.

\begin{figure}
\includegraphics[width=8cm]{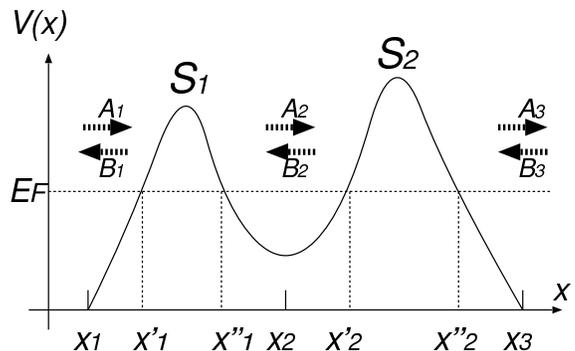}
\caption{Potential profile for the effective model of resonance
tunnelling. Heights of the two potential peaks are characterized
by $S_1$ and $S_2$.} \label{fig:potential}\end{figure}

Here we describe the resonance-tunnelling theory, following
Ref.~\onlinecite{Azbel1983}, and will see whether it fits with the
present model with disorder. We define an effective model of
``resonance tunnelling'' as described by the Schr{\"o}dinger
equation $\Psi''(x)+(k^{2}-V(x))\Psi(x)=0$ where
$\hbar^{2}k^{2}/2m$ is the particle energy, and the effective
potential $V(x)$ has two peaks as shown in
Fig.~\ref{fig:potential}. It is not trivial whether the numerical
results of our model (\ref{eq:H_pure}) fits well with this
resonance-tunnelling picture. The present model has many valley
and peaks, due to on-site randomness, and is not similar with
Fig.~\ref{fig:potential}. Nevertheless, when the localization
length $\xi$ is much shorter than the system size, it is indeed
the case as we see by fitting our numerical results well by the
picture of resonance tunnelling. The basic reason for the
applicability of resonance tunnelling is that the topology
guarantees an existence of a perfect-transmittance point, i.e.,
the resonant tunnelling in the two-dimensional parameter space
$(Q_1,Q_2)$, in spite of the complexity of the potential shape in
the model.

First let us consider an open system attached to two ideal leads
which are connected to reservoirs. The Fermi energy $E_F$ is tuned
at one's disposal, and we assume that the potential peaks are much
higher than $E_F$. An effective wave number $k(x)\equiv
\sqrt{k^{2}-V(x)}$ becomes imaginary inside the potential peaks,
and the particle tunnels through the peaks. One of the eigenstates
at the left lead ($x<x_1$) is given by $\Psi_1(x)\sim
\exp(ik_1(x-x_1))$, where $k_{1}=k(x_{1})$. The time-reversal
symmetry of the system requires that the complex conjugate
${\Psi_1}^{*}(x)$ is also an eigenstate. At the potential valley
around $x=x_2$ we can also consider eigenstates with
$\Psi_2(x)\sim \exp(ik_2 (x-x_2))$ and ${\Psi_2}^{*}(x)$, where
$k_{2}=k(x_{2})$. Thus general eigenstates around $x=x_1$ and
around $x=x_2$ are given as
$\Psi(x)=A_1\Psi_1(x)+B_1{\Psi_1}^{*}(x)$ and
$\Psi(x)=A_2\Psi_2(x)+B_2{\Psi_2}^{*}(x)$, respectively. Next we
introduce the transfer matrix $\Theta_{1}$ from $x_2$ to $x_1$
\begin{equation}
\left(\begin{array}{c}
A_1\\B_1
\end{array}\right)
=\Theta_{1}
\left(\begin{array}{c}
A_2\\B_2
\end{array}\right),
%\ \ \Theta=\left(
%\begin{array}{cc}\Theta_{11} &\Theta_{12}\\\Theta_{21} &\Theta_{22}
%\end{array}
%\right).
\end{equation}
where $\Theta_{1}$ is a $2\times 2$ matrix.
Time-reversal symmetry requires the transfer matrix $\Theta$ to be of
the form

\begin{equation}
\Theta_{1}=\left(\frac{k_{2}}{k_{1}}\right)^{\frac{1}{2}}\left(
\begin{array}{cc}
\cosh (S_1)e^{i\alpha_1}&
\sinh (S_1)e^{i\beta_1}\\
\sinh (S_1)e^{-i\beta_1}&
\cosh(S_1)e^{-i\alpha_1}\\
\end{array}
\right).
\end{equation}
In a semiclassical theory we have
\begin{eqnarray}
&&\alpha_{1}=-
\int_{x_{1}}^{x'_{1}}k(x)dx-\int_{x''_{1}}^{x_{2}}k(x)dx, \\
&&\beta_{1}=\frac{\pi}{2}-
\int_{x_{1}}^{x'_{1}}k(x)dx+\int_{x''_{1}}^{x_{2}}k(x)dx.
\end{eqnarray}
$S_1$ characterizes the height of the potential, as the transmission
coefficient $t_1$ is given by $t_1=e^{\-i\alpha}
\mathrm{sech}S_1$. Because
we assume that the peak is high, which means $S_1\gg 1$.
The transmission probability $|t_1|^{2}$ is then approximated as
$|t_1|^{2}\sim 4e^{-2 S_1}\ll 1$.

Similarly, we introduce the transfer matrix for the second peak as
\begin{equation}
\left(\begin{array}{c}
A_2\\B_2
\end{array}\right)
=\Theta_{2}
\left(\begin{array}{c}
A_3\\B_3
\end{array}\right),
\end{equation}
and we have
\begin{equation}
\Theta_{2}=\left(\frac{k_{3}}{k_{2}}\right)^{\frac{1}{2}}\left(
\begin{array}{cc}
\cosh(S_2)e^{i\alpha_2}&
\sinh(S_2)e^{i\beta_2}\\
\sinh(S_2)e^{-i\beta_2}&
\cosh(S_2)e^{-i\alpha_2}\\
\end{array}
\right).
\end{equation}
The phases $\alpha_{2}$ and $\beta_{2}$ can be written
similarly to $\alpha_{1}$ and $\beta_{1}$, where $S_{2}\gg 1$.
We assume that $S_{i}$, $\alpha_{i}$, and $\beta_{i}$ $(i=1,2)$ are
smooth functions of the parameters $Q_1$, $Q_2$ and $E_F$.
A transfer matrix $\Theta$ for the entire system is given by
\begin{eqnarray}
&&\Theta=\Theta_{1}\Theta_{2}
=\left(\frac{k_{3}}{k_{1}}\right)^{\frac{1}{2}}\left(
\begin{array}{cc}
\theta_{11}&\theta_{21}^{*}\\
\theta_{21}&\theta_{11}^{*}
\end{array}
\right),\\
&& \theta_{11}=e^{i(\alpha_{1}+\alpha_{2})}
\left(\cosh S_{1}\cosh S_{2}+\sinh S_{1}\sinh S_{2}e^{i\omega}
\right),\nonumber \\
&& \\
&& \theta_{21}=e^{i(\alpha_{2}-\beta_{1})}
\left(\sinh S_{1}\cosh S_{2}+\cosh S_{1}\sinh S_{2}e^{i\omega}
\right),\nonumber \\
&&\\
&&\ \ \ \ \ \omega=\beta_{1}-\beta_{2}-\alpha_{1}-\alpha_{2}\cong
2\int_{x''_{1}}^{x'_{2}}k(x)dx.
\end{eqnarray}
>From unitarity, it follows that $|\theta_{11}|^{2}-|\theta_{21}|^{2}=1$.
We assume that the Fermi energies of the two leads are identical,
which means $k_{3}=k_{1}$. For a plane wave incident from the left lead,
let $r$ and $t$ denote
the reflection and transmission coefficients,
respectively. Similarly, for a plane wave from the right lead, we
define $r'$ and $t'$ as well. We then obtain
\begin{equation}
r=\theta_{21}/\theta_{11},\ t=t'=1/\theta_{11}, \
r'=-\theta_{21}^{*}/\theta_{11}.
\end{equation}
It implies the unitarity $|r|^{2}=|r'|^{2}=1-|t|^2$.
It also satisfies $t=t'$ and $rt^{*}+r'^{*}t=0$, as is required
from time-reversal symmetry.

\begin{figure}[htb]
\includegraphics[width=8.5cm]{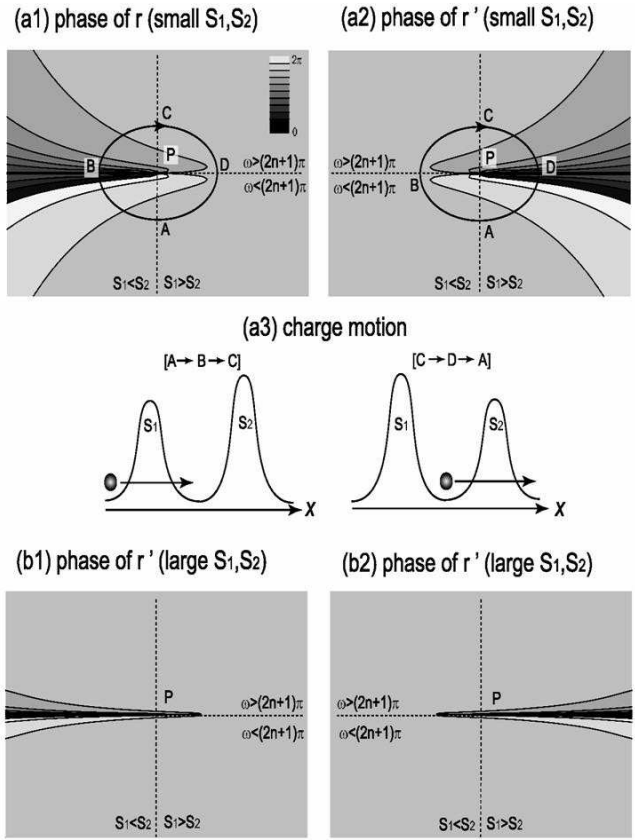}
\caption{Schematic contour mapping of the
phases of the reflection coefficients
$r$ and $r'$ in the $Q_1$-$Q_2$ plane.
(a1)(a2) correspond to small $S_1$, $S_2$, and (b1)(b2) to large $S_1$,
$S_2$.
The phases of $r$ and $r'$
wind by $2\pi$ during the cycle around the ``vortex'' $P$.
When we go along the arrows indicated in the figures (a1)(a2), the phase of
$r$ ($r'$)
increase by $2\pi$ ($-2\pi$).
(a3) illustrates the motion of the charge in the cycle.
Note that the $2\pi$ phase change of $r$ ($r'$)
is associated with
pumping one unit charge into the system through the left (right) end of the
system.}
\label{fig:resonant}\end{figure}

We now apply this framework to fit our numerical results.
Because $r$ is written as
\begin{equation}
r=e^{i\theta}
\frac{\tanh S_{1}+\tanh S_{2}e^{i\omega}}{1+\tanh S_{1}
\tanh S_{2}e^{i\omega}},
\end{equation}
a condition for a total transmission, $r=0$, is given by
\begin{eqnarray}
&&S_{1}=S_{2},\label{eq:S1S2}\\
&&\omega= 2\int_{x''_{1}}^{x'_{2}}k(x)dx
=(2n+1)\pi \label{eq:omega},
\end{eqnarray}
where $n$ is an integer. The latter condition is
equivalent to the Bohr quantization condition, that the state localized
around $x_2$ be an eigenstate with its
eigenenergy equal to $E_F$.
In other words the total transmission occurs at resonance.
For fixed $E_F$, the two conditions (\ref{eq:S1S2}) and (\ref{eq:omega})
define isolated points in the $Q_1$-$Q_2$ plane.
Let $P$ denote one of such points of total transmission: $r=0$.
One can easily see that around the point $P$, the phase of $r$ and
that of $r'$ wind by $\pm 2\pi$, as is schematically
shown in Fig.~\ref{fig:resonant}(a1)
(a2).
As is seen from the Brouwer formula (\ref{eq:q}), the phase windings
of $r$ and $r'$ correspond to ($2\pi$ times) the amount of charge
pumped into the system through the left and the right ends, respectively.
Thus, by going around the point $P$, one unit charge is pumped from
the left lead to the right.
For illustration, let us consider
a clockwise cycle around the point $P$ in Fig.~\ref{fig:resonant} (a1)(a2).
This pumping from left to right is analogous to the ``bicycle pump''
~\cite{Avron2004}.
The two potential peaks correspond to two gates to control the pumping.
If the system crosses the line $\omega=(2n+1)\pi$ at the $S_1<S_2$ side,
$r$ undergoes $2\pi$ phase changes. It
means that tunneling occurs through the left potential peak due to resonance,
corresponding to
the opening of a ``left gate'' and  a unit charge flows in.
After that the left gate closes, and the right gate opens in turn,
when the system crosses the line $\omega=(2n+1)\pi$ at the $S_1<S_2$ side,
and $r'$ undergoes $-2\pi$ phase change. The charge is pumped to the
right lead after one cycle.
Thus the overall motion of the charge is as shown in
Fig.~\ref{fig:resonant} (a3).
For the pumped charge to be quantized, the cyclic process should
be sufficiently slow to be regarded as ``adiabatic''. Otherwise
the particle cannot tunnel through the potential barriers.
Below let us make this physical picture more explicit in relation to
our numerical results.

In our numerical results in Fig.~\ref{fig:r-phase}(b2)-(b4), the
phase-winding point $P$ corresponds to the perfect transmission.
When we go from Fig.~\ref{fig:r-phase}(b2) to (b4), $L/\xi$
increases. In the resonance-tunnelling picture, it means that
$S_i$ increases and that the ``resonance-tunnelling wing''
narrows, as is seen by comparing Fig.~\ref{fig:resonant}(a1) and
(b1). It is exactly observed in our numerical results.

For large $L/\xi$, the transmission probability $|t|^{2}$ is
typically $\sim e^{-2(S_1+S_2)}\sim e^{-2L/\xi}$. Because we have
$S_1\sim S_2$ near the resonance, we estimate $S_1\sim S_2\sim
(L/2\xi)$. In that case, the change of the phases of $r$ and $r'$
occurs abruptly at around $|\omega - (2n+1) \pi| \sim e^{-
L/\xi}$. Along the line $\omega=(2n+1) \pi$, $|r|$ is given by
$|r|=\tanh (S_1-S_2)$, which becomes  appreciable only when
$|S_1-S_2| \sim 1$. This is found only in a small region
$\Delta/t_{\mathrm{n.n.}} \sim \xi/L$. When $L/\xi\gg 1$, because
the tunnelling rate $Q_W$ is exponentially small ($Q_W \sim
t_{\mathrm{n.n.}} e^{-L/\xi} $), charge pumping requires an
exponentially long time $\tau \sim (\hbar/t_{\mathrm{n.n.}})
e^{L/\xi}$. This is a time scale which gives a criterion for
adiabatic charge pumping in this system. If one changes the
parameters faster than this time scale, the pumped charge is
reduced with an exponential factor $e^{-\omega/E_G}$ as a
nonadiabatic correction, where $\omega$ is a frequency of the
change the external parameters~\cite{ShihNiu94}.

Generally, there occurs no perfect transmittance point by tuning
only two parameters since the effective two barrier model for the
resonant tunnelling applies only to a limited region of the random
system and not through a whole sample. In the present case it is
guaranteed by the topological constraint that the phase of $r$
($r'$) should wind by $(-)2 \pi$ for a large cycle far from
$\vec{Q}=0$, well within the gapped region~\cite{NiuThouless84}.
When $|{\vec Q}|$ is larger than the energy scale of the disorder,
a finite gap ($\sim |\vec{Q}|$) opens. The wing will then become
as wide as $Q_W\sim|\vec{Q}|$. Correspondingly, the typical time
scale $\tau\sim \hbar/Q_W \sim \hbar/|\vec{Q}|$ becomes smaller,
and the adiabaticity condition is easily satisfied. In such case,
however, the dielectric response is not enhanced.
%{\bf{ Here we should stress that even in the strongly localized case, i.e.,
%disorder $\gg$
%transfer, one can satisfy the adiabatic condition very easily when
%$Q_1,Q_2$ are large enough and there remains a finite gap. In this case,
%the resonant tunneling state behaves like an extended state for the
%purpose of charge pumping. The trick is that the parameter sets
%of the resonant tunneling state is never realized and only acts as
%a fictitious point.}}
The localization length becomes as long as the system size, and
the pumping is accomplished through extended states, not by
resonance tunnelling. Remarkably, this charge pumping in the
gapped region is governed by the vortex which is located deep in
the disordered regime ($\xi\ll L$). In other words, the charge
pumping in the gapped regime ($\xi\sim L$) is smoothly connected
to that via resonance tunnelling in the disordered regime ($\xi\ll
L$). We note that $S_1$ and $S_2$ are regarded as effective
parameters, although the real potential is much more complicated
than the two-barrier structure.
%$S_1$ and $S_2$ are more directly relevant to double-dot
%systems~\cite{double-dot}.

\subsection{Multi-channel problem in alloy model for on-site random potential}

In the rest of this section, we consider a stronger randomness
by employing the alloy model (Fig.~\ref{fig:random} (b)) with $v_i= s_i(v+\delta v_i)$
with $s_i=\pm1$ being a random sign and
$\delta v_i/v\in[-0.025,0.025]$ a uniform random distribution, instead of the uniformly distributed random potential (Fig.~\ref{fig:random} (a)).

\begin{figure}[htb]
  \begin{center}
    \includegraphics[width=8cm]{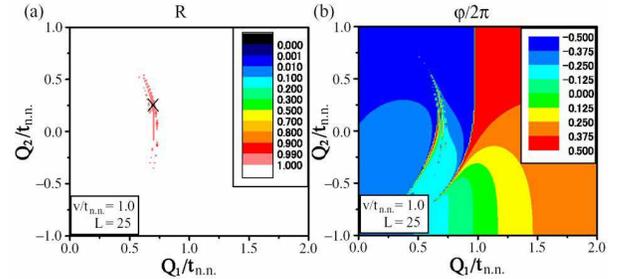}
  \end{center}
  \caption{(a) Reflectance and (b) the phase $\varphi$ of $r$
    in disordered system with $L=25$ and $v/t_{n.n.}=1.0$ for the alloy model with the on-site random potential given by Fig.~\ref{fig:random}.}
  \label{fig:r-phase_alloy}
\end{figure}

Let us start with the single-channel case. Then, for a fixed
disorder strength $v$, the shape of the vortex core of the
reflection coefficient rapidly evolves from isotropic to
anisotropic with increasing the system size $L$, as in the case of
the uniformly distributed random potential discussed in
Sec~\ref{subsec:single}. This tendency is generic and also
realized for a fixed system size $L$ with increasing the disorder
strength $v$. In Figs.~\ref{fig:r-phase_alloy} (a) and (b), we
show the reflectance $R$ and the phase $\varphi$ in the $\vec{Q}$
space for $v/t_{\mathrm{n.n.}}=1.0$ and a small sample size
$L=25$. Here, the region in the $\vec{Q}$ plane where the gap
collapses and vortices distribute expands because of the stronger
disorder and thus the much shorter localization length $\xi \sim3$
or 4. The transmittance $T$ is typically obtained as
$\sim10^{-9}$, which is practically negligible except at the
vortex core.

\begin{figure}[htb]
  \begin{center}
    \includegraphics[width=8cm]{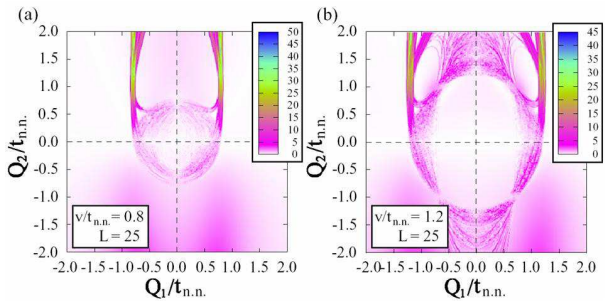}
  \end{center}
  \caption{Relative dielectric response in the presence of the disorder compared with the pure case. The disorder strength is $v/t_{n.n.}=0.8$ for (a) and $1.2$ for (b). Averages are taken over $10^2$ random disorder configurations. Inside white bands, there occurs a gradual sign change in the dielectric response.}
\label{fig:enhance-open}
\end{figure}

Now we consider a system that consists of many channels, each of
which is described by the present disordered alloy model but with
a different profile of random potentials. Such configuration can
be realized in thin films of ferroelectrics. Then, we can design
the pattern of the phase $\varphi$ in the $\vec{Q}$ plane by
tuning the disorder. For the alloy model with the random on-site
potentials, vortices are mostly located around $Q_1=\pm v/2$ with
the ``wing'' almost along the $Q_2$ axis. Therefore, we can
enhance the dielectric response, if we can tune the disorder
strength and choose a sample where the $\vec{Q}$ point of the
system is located inside the ``wing''. In particular, when the
transmittance $T$ is negligibly small, the enhancement factor of
the dielectric response is given by
$(\partial\varphi/\partial\theta)_{\rm
disorder}/(\partial\varphi/\partial\theta)_{\rm pure}$ from
(\ref{eq:q}), since the electric field is proportional to
$\theta=\arctan(Q_1/Q_2)$. We calculate this enhancement factor
for this multichannel system. Increasing the number $N$ of
channels. Figures~\ref{fig:enhance-open}(a) and (b) represent the
$\vec{Q}$ dependence of the enhancement factor in the case of
$v/t_{n.n.}=1.6$ and $2.4$, respectively, for $L=25$ with $N=102$.
The main structure in this map is almost saturated up to $N=10^2$.
These results reveal that around $Q_1\sim\pm v/2$, the dielectric
response is significantly enhanced by a factor $30\sim40$ compared
with the pure case. Even for the thin film with a square shape of
a linear dimension larger than $50 \mbox{\AA}$, which corresponds
to $N=10^2$, the disorder-induced enhancement of the charge
transfer rate should be robust. Then, the applied electric field
necessary for switching the polarization is reduced by this
enhancement factor. If we require the response time $\tau\sim
e^{L/\xi}/t_{\mathrm{n.n.}}$ of the order of $10^{-9}\ {\rm s}$,
we obtain $e^{L/\xi}<10^6$ with the assumption of
$t_{\mathrm{n.n.}}\sim 10^{15}\ {\rm s}^{-1}$. This also assures a
negligibly small transmittance $T\sim e^{-L/\xi}\sim 10^{-6}$, and
thus the small leak current and low dissipation.

Possible experimental realization of this quantum-mechanical disorder-induced enhancement of the dielectric response, namely the quantum relaxor, has also been proposed~\cite{onoda2006prl} for thin films of solid solution systems like Pb(Fe$_{0.5}$Nb$_{0.5}$)O$_3$ and Pb(Sc$_{0.5}$Nb$_{0.5}$)O$_3$ prepared with an adequate slow-anneal process~\cite{Galasso,AsanumaJJAP}.

\section{Periodic/Twisted boundary condition}
\begin{figure}[htbp]
  \includegraphics[scale=0.4]{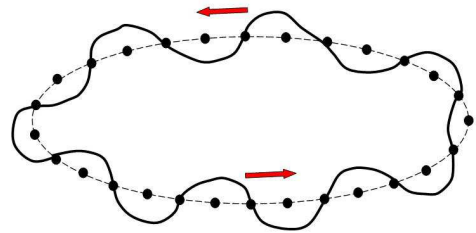}
  \caption{\label{fig:1dring} A cartoon picture of the charge transport on the one-dimensional ring.}
\end{figure}

The adiabatic charge pumping in the presence of the substrate
disorder was considered by Niu \emph{et al.}\cite{NiuThouless84}.
They showed that the adiabatic charge transport is still quantized
as long as the excitation gap between the highest occupied state
(HOMO) and lowest unoccupied state (LUMO) does not vanish due to
the substrate disorder and the many-body interaction in the
thermodynamic limit. In the case of open boundary condition, it is
easy to imagine the meaning of the charge pumping, namely the
charge transport from one end to the other. In the case of
periodic boundary condition, the charge pumping after one-cycle
means that the electronic wavefunction shifts by the minimum
number of lattice sites so that the final wavefunction is the same
as the initial one.  A cartoon picture of this process is given by
Fig.~\ref{fig:1dring}.
We note that in the absence of the disorder, the
adiabatic charge transport has been related to the
field-theoretical model of the one-dimensional chiral
anomaly\cite{zee1984}.
%One of our main result shown in the
%following is that the chiral anomaly can be destroyed by strong
%random disorder, where the energy gap is the order of band width
%divided by the size of the system.

\subsection{Formalism}\label{section:formalism}
We consider a one-dimensional charge-density-wave system
with disorder, as described in (3) and (7).
In contrast to the open system with two leads as considered in
Section III, we deal with a closed system in a ring geometry.
Instead of a periodic boundary condition as is usually employed for the
ring, we introduce a twisted boundary condition;
we impose that the phase gained by running around the ring once is
$e^{i\alpha}$.
%where $N$ is the number of sites in the ring.
This twisted boundary condition
is realized by
applying a magnetic flux through the one-dimensional ring.

The reason for introducing the twisted boundary condition is to
reveal the topological nature of the charge transport in
the system.
In the pure case, the lattice momentum $k$ is a good quantum
number,
and the topological nature is manifest in the three-dimensional
space $(Q_1,Q_2,k)$ as monopoles and antimonopoles.
However, $k$ is no longer a good quantum number
in the disordered case, and instead
the flux $\alpha$ plays a similar role to $k$.
%The quantization of
%the magnetic flux is synonymous with the discretized lattice
%momentum.
Therefore, instead of $(Q_1, Q_2, k)$, we will be
interested in the adiabatic process such that the system is
changed due to the slow variations of the parameters $(Q_1, Q_2,
\alpha)$, which spans a 3-dimensional parameter space with the
property $\alpha\in [0, 2\pi]$.
The Hamiltonian with the flux $\alpha$ can be
rewritten as
\begin{eqnarray}
H(\alpha)=&-&\frac{t_{\text{n.n.}}}{2}
\sum_{j=1}^N (e^{i\alpha/N}c^\dag_jc_{j+1}+h.c.)\nonumber \\
&
+&
Q_1\sum_{j=1}^N (-1)^jc^\dag_jc_j \nonumber \\
&+& \frac{Q_2}{2} \sum_{j=1}^N (-1)^j(e^{i\alpha/N}c^\dag_jc_{j+1}+h.c.)+V,
\label{2-delta-hamiltonian-with-phi}
\end{eqnarray}
where $c_{N+1}\equiv c_{1}$ and
$V=\sum_j v_j c^\dag_j c_j$ is the uniformly
distributed disorder potential.

In the tight binding model, the polarization operator $\vec{P}$
has the following form
\begin{eqnarray}
\vec{P}=\sum_j \vec{R}_jc^\dag_jc_j
\end{eqnarray}
where $\vec{R}_j$ is the position at site $j$ and $c^\dag_jc_j$ is
the electron density operator.  The current is defined as the time
derivative of the polarization operator given by
\begin{eqnarray}
\vec{J}=\frac{\partial\vec{P}}{\partial t}=\frac{1}{i}[\vec{P},H]
\label{current}
\end{eqnarray}
In the pure case the current is given by $\frac{\partial H}{\partial
k}$,
whereas in the present case with disorder,
the current operator given by $\frac{\partial H}{\partial \alpha}$
has the following form
\begin{eqnarray}
J=ie^{-i\alpha/N}\sum_j(t+Q_2(-1)^j)c^\dag_{j+1}c_j + h.c.
\end{eqnarray}

When the flux $\alpha$ is equal to $0$ or $\pi$, the system
is time-reversal symmetric, and there is no persistent current.
In such cases, we can consider a change of the electric polarization
by an adiabatic change of parameters $Q_1$ and $Q_2$.
In the linear response theory, the change of the electric
polarization is given by
\begin{equation}
\delta P=F_{1}\Delta Q_{2}-F_{2}\Delta Q_{1},
\label{deltaP}
\end{equation}
with
\begin{eqnarray}
F_{1}
%\frac{\partial P}{\partial Q_2}
=-\frac{i}{L}\sum_{m\neq 0}
(\frac{\langle\Psi_0|J|\Psi_m\rangle\langle\Psi_m|\frac{\partial H}{\partial
Q_2}|\Psi_0\rangle}{(E_m-E_0)^2}-c.c.), \label{dpolar_dD2}\\
F_{2}=
%\frac{\partial P}{\partial Q_1}=
\frac{i}{L}\sum_{m\neq 0}
(\frac{\langle\Psi_0|J|\Psi_m\rangle\langle\Psi_m|\frac{\partial H}{\partial
Q_1}|\Psi_0\rangle}{(E_m-E_0)^2}-c.c.), \label{dpolar_dD1}
\end{eqnarray}
where $L=Na$ is the circumference of the ring, and $a$ is the
lattice spacing, and
\begin{eqnarray}
& &\frac{\partial H}{\partial Q_1}=\sum_j(-1)^jc^\dag_jc_j \\
& &\frac{\partial H}{\partial
Q_2}=\frac{1}{2}
\sum_j(-1)^j(e^{i\alpha/N}c^\dag_jc_{j+1}+e^{-i\alpha/N}c^\dag_{j+1}c_j).
\end{eqnarray}
We note that the change of the polarization, $\delta P$, is
in general dependent on the path in the parameter ($\vec{Q}$) space.
%Note that $\frac{\partial P}{\partial \alpha}$ is manifestly
%vanishing.
The $|\Psi_0\rangle$ in (\ref{dpolar_dD2}) and
(\ref{dpolar_dD1}) is the many-body ground state, and $|\Psi_m\rangle$
denote the excited states.  $E_0$ and $E_m$ are the energy for the
ground and the excited states, respectively.  In the pure case
when the chemical potential lies in the gap, the system is a band
insulator. The magnitude of the energy gap is given by the magnitude
of the CDW order parameter $(Q_1, Q_2)$.  In the presence
of the disorder, the gap closes when the magnitude of the disorder
potential becomes the order of $\mathcal{O}(\sqrt{Q_1^2+Q_2^2})$.
The system
remains insulating because each state is localized, i.e., Anderson
localization.

Although we have defined
(\ref{deltaP}), (\ref{dpolar_dD1}) and (\ref{dpolar_dD2})
only for $\alpha=0$ and $\alpha=\pi$, we henceforth extend these
formulae to general $\alpha$, which
makes the topological properties of the charge pumping manifest.
We note that except for $\alpha=0$ and $\pi$, $\delta P(\alpha)$ does not
mean a change of polarization.
%In fact, both (\ref{dpolar_dD1}) and (\ref{dpolar_dD2}) have
%geometrical meanings.
With this extension to arbitrary $\alpha$, let us succinctly
abbreviate $(Q_1,Q_2,\alpha)$ as ${\vec Q} =(Q_1,Q_2,Q_3)$ and
define the gauge potential as
\begin{eqnarray}
\vec{A}&=&i\langle\Psi_0|\frac{\partial}{\partial
\vec{Q}}|\Psi_0\rangle \label{gauge_potential}.
\end{eqnarray}
This gauge field is so defined that the
corresponding field strength $\vec{F}=\nabla\times\vec{A}$
has the components given in
(\ref{dpolar_dD1}) and (\ref{dpolar_dD2}).
%are then the
%components of the field strength
%\begin{eqnarray}
%F_1=\frac{\partial P}{\partial Q_2}, \ F_2=-\frac{\partial
%P}{\partial Q_1}.
%\end{eqnarray}
Furthermore,
when the parameters $(Q_1,Q_2)$ are changed along a cycle,
the pumped charge  can be written as
\begin{eqnarray}
\delta P(\alpha) = \oint_S
d\vec{Q}\times\hat{\alpha}\cdot\vec{F}(\vec{Q}) \label{gauge_flux}
\end{eqnarray}
where $S$ is a loop on the $Q_1$-$Q_2$ plane.
Using the
Stoke's theorem, (\ref{gauge_flux}) becomes
\begin{eqnarray}
\delta P(\alpha) =\int d^2 Q D(\vec{Q})
\label{distribution}
\end{eqnarray}
where
$D(\vec{Q})=\partial_1 F_1+\partial_2 F_2$
%\vec{\nabla}\times(\hat{\alpha}\times\vec{F})
defined as the distribution function of the polarization.
This pumped charge is not necessarily an integer.

Nevertheless, in a strongly disordered system it becomes an
integer, with an exponentially small
correction of the order of $e^{-L/\xi}$.  To
see this, we note that the wavefunctions in a strongly localized case
are almost intact with the twisted boundary condition $\alpha$
within an accuracy of $e^{-L/\xi}$.
Hence $\Delta P(\alpha)$ is equal to
an average $\Delta P(\alpha)$
over $\alpha$,
namely
\begin{eqnarray}
\Delta P = \int_0^{2\pi}\frac{d\alpha}{2\pi}\Delta
P(\alpha)=\int_{\partial T} d\vec{\sigma}\cdot \vec{F}
\label{torus_surface}
\end{eqnarray}
where $\partial T$ is the torus surface, and $d\vec{\sigma}$ is
the surface element. This average $\Delta P$ represents the total
flux over a torus surface, and it is an integer, which follows
from topological properties of the gauge field $\vec{A}$. As the
field strength is defined as $\vec{F}=\nabla\times\vec{A}$, it
satisfies $\nabla\cdot\vec{F} =0$ \textit{if there is no
singularity, i.e. there is no energy crossing}. Thus when there
are no energy degeneracies inside the torus $V\times [0,2\pi]$,
the averaged pumped charge $\Delta P$ becomes zero due to the
Gauss theorem. If there is a energy crossing, the gauge field
$\vec{A}$ has a U(1) monopole there, and the averaged charge
$\Delta P$ becomes a total strength of the monopoles inside the
torus, again by using the Gauss theorem. As the monopole strength
is quantized to be an integer, $\Delta P$ is always an integer.
This discussion for (\ref{torus_surface}) corresponds to that for
(2.23)in Ref. \onlinecite{NiuThouless84}, in which they claimed
that the charge transport is quantized in the presence of gap
which does not vanish in the thermodynamic limit. On the other
hand, the aim of this paper targets at another respect: we are
interested in the strong disorder case, in which the energy gap
closes in the thermodynamic limit and all wave functions are
localized characterized by the localization length $\xi$ defined
by $\xi^{-1}=\int dx |\psi(x)|^4$, where $\psi$ is the single
particle wave function. Nevertheless, for finite-sized systems gap
always exists which is the order of the band width divided by the
size of the system, which validates the above topological argument
even in the strongly disordered cases. In such disordered systems,
the field strength $\vec{F}$ is highly anisotropic around the
monopoles, which is closely related with a ``resonance
tunnelling'' picture in section III.B as we explain in the next
paragraph.

The adiabatic charge transfer is based on the occurrence of the
singularity in the parameter space\cite{OnodaMurakamiNagaosa04},
which happens when LUMO and HOMO are degenerate.  By making a
close contour enclosing the singularity, the wave function shifts
as shown in Fig.~\ref{fig:1dring}.  Since the disorder potential
is random, we expect that the proliferation of the singularities
may occur. Furthermore, there could be \emph{singularity string}
rather than \emph{point} in the strong disorder limit. The reason
is given as the following.  In the strong disorder limit, all wave
functions are localized. There could be a situation that the wave
function of the LUMO and the HOMO are well separated so that the
overlap integral of two wavefunctions are
%essentially zero
exponentially small
as the
parameters change.
Thus even when the
change of parameters in numerics or in experiments is very slow,
its timescale might be still shorter than the inverse of the
(exponentially small) overlap integral of the LUMO and HOMO.
Therefore, within this variation of parameters, the LUMO and
HOMO can be regarded as degenerate.
Once they are degenerate at a
certain point in the parameter space, they will keep the
near-degeneracy as the parameters change until their wave functions
have
%finite
appreciable overlap
%so that the off-diagonal matrix elements
%between them to be non-zero
and open a gap. As long as they remain nearly degenerate as the
parameters change, the singularity string is created in the
parameter space. Moreover, because the wave functions are highly
localized on the singularity string, they are \emph{insensitive to
the boundary condition}, in other words, $\alpha$-independent.
Therefore, they have the sheet-like structure extending along the
$\alpha$-direction in the parameter space. Even though it is
sheet-like, we call it a string referring to its projection on the
$(Q_1, Q_2)$ plane.  From this observation, we conclude that the
singularity in the parameter space projected on $Q_1$ and $Q_2$
subspace has two kinds, points and strings, in the strong disorder
limit.  The former contribute to the charge transfer, while the
latter do not. The charge transfer for the former is explained
within the resonance tunnelling
%In the second kind,
%the wave functions of LUMO and HOMO do not overlap. In the first
%kind, their wave functions overlap and lead to resonance
%tunneling which will be explained in the section
Section \ref{section:resonance-periodic}.

\subsection{Numerical Result}\label{section:numerical}
In this section, we solve numerically the eigenvalue problem for
slowly varying parameters $(Q_1,Q_2)$ and calculate the contour
integral in (\ref{gauge_flux}).
We introduce the small segment $\Delta {\vec Q}$ for
the contour integral, and demand that $F_1,F_2$ are
continuous. Therefore, the size $|\Delta{\vec Q}|$ determines the
energy-scale and time-scale of our simulation. In actual
calculation, we take
$|\Delta{\vec Q}|=10^{-6}t_{\text{n.n.}}$ as the finest one.
Then the discussion below applies when we observe the system
within the time-scale $T \sim \hbar/|\Delta{\vec Q}| \sim 10^6 \hbar/t_{\text{n.n.}}
$.
Also we average over $\alpha$ to see the quantization of the
charge transfer in (\ref{torus_surface}).

We solve (\ref{2-delta-hamiltonian-with-phi}) by a direct
numerical calculation.  Our goal is to compute the distribution
function from the (\ref{distribution}), which is
$\alpha$-dependent. We will also demonstrate the quantization of
the charge transfer by averaging over $\alpha$ by
(\ref{torus_surface}).

We consider a ring with $N=50$ and take $v_i/t_{\text{n.n.}}$ to distributes
uniformly in $[-7.0,7.0]$.  To obtain the distribution function,
we divide the $(Q_1, Q_2)$ plane into small pieces of
grids and calculate (\ref{distribution}) for each grid.  Let us
remind the readers that the distribution function for the pure
case is the $\delta-$function at the
origin\cite{OnodaMurakamiNagaosa04}. In that case, we do not need
to integrate over $\alpha$ to obtain the quantization of the
charge transfer, because the singularity lives in the $(Q_1,
Q_2, k)$ space, and the integration over $k$ is already
included implicitly in (\ref{distribution}).

\begin{figure}[htbp]
  \includegraphics[scale=0.4]{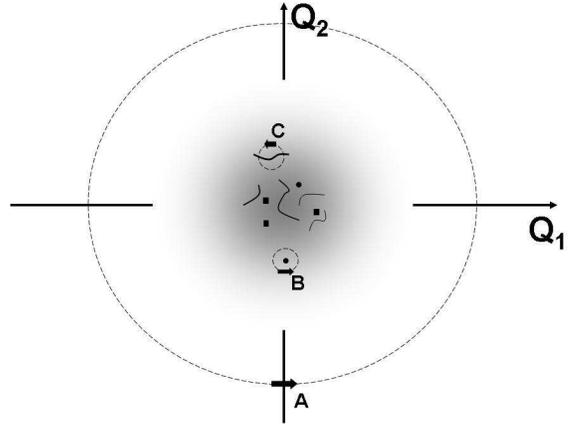}
  \caption{\label{fig:distribution} A cartoon picture of the distribution
of singularities
at $\alpha = 0$.  The singularity has two kinds: strings and points.
  Calculating (\ref{gauge_flux}) from the outer big dash loop gives the polarization +1.0 and -1.0 from the loop {\bf B}.  If we stay far away from
  the origin, one would think that there is only one monopole with $\Delta P = 1.0$ at the center like the pure case.  However, when the disorder strength
  is strong enough, there are some subtle structure inside the big monopole.  There are the monopole-antimonopole pairs and the singularity strings.}
\end{figure}

\begin{figure}[htbp]

    \includegraphics[scale=0.4]{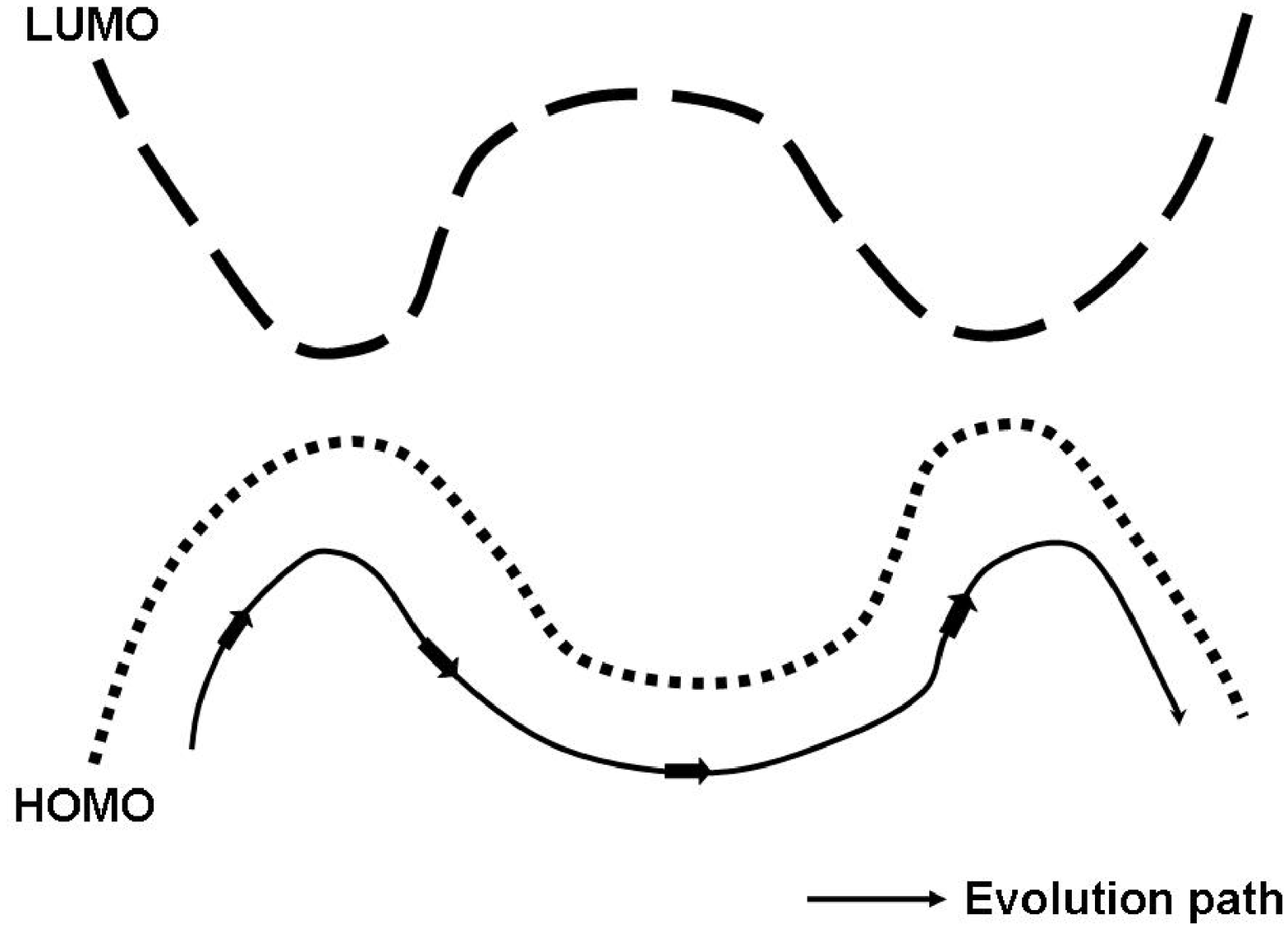}
    (a)

    \includegraphics[scale=0.4]{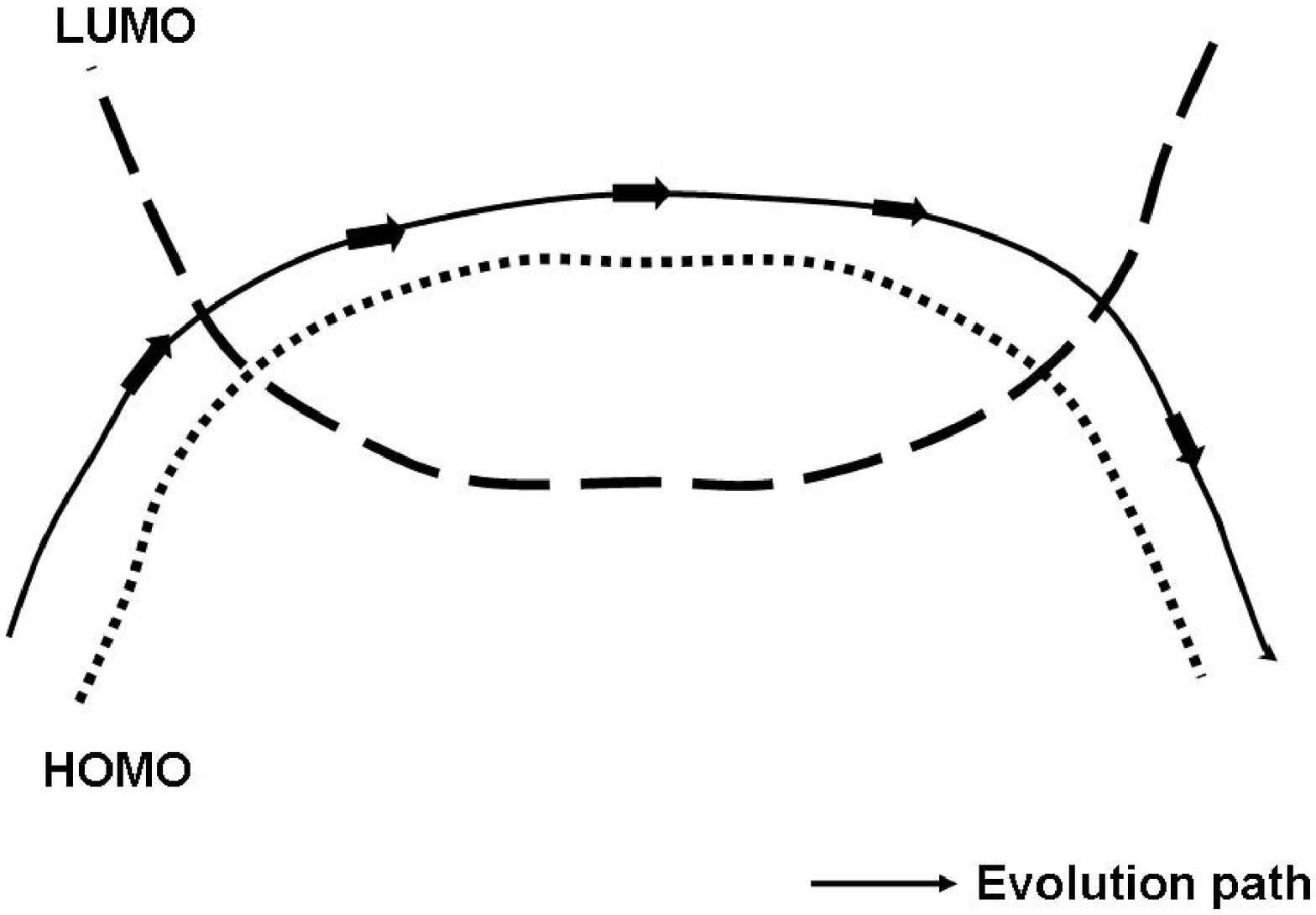}
    (b)

    \includegraphics[scale=0.4]{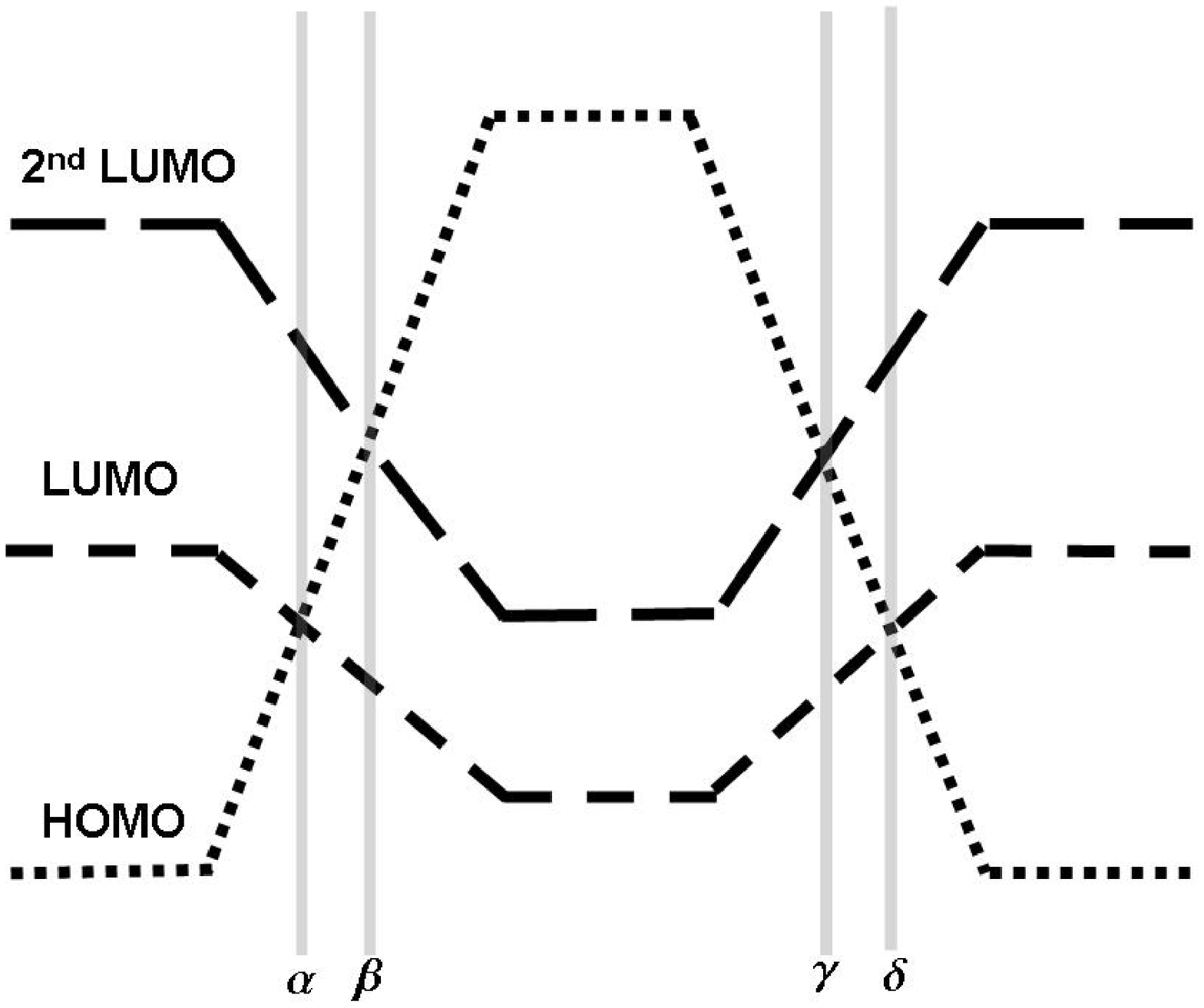}
    (c)

\caption{\label{total_adiabatic} The adiabatic process in the
gapless case is different from the gap case.  (a) shows the
process with a gap.  (b) shows the process without a gap.  (c) a
real process from loop {\bf C} in the
Fig.~\ref{fig:distribution}.}
\end{figure}

Our result can be summarized by the cartoon in
Fig.~\ref{fig:distribution}. First, we found that $\Delta P=1.0$
if we take the loop {\bf A} which is the outer dash one ($|{\vec
Q}| \sim 10$) and average over $\alpha$.  It corresponds to the
weak disorder case where the strength of the disorder is much
smaller than $|\vec{Q}|$, so the potential strength is not large
enough to close the gap.  According to Niu and
Thouless\cite{NiuThouless84}, the polarization is quantized, and
we reproduce their result here. Secondly, the singularity strings
appear in the region $|{\vec Q}| \leq 4.5$. In this region, the
disorder potential is strong enough to close the gap. As we
pointed out in the section \ref{section:formalism}, there are two
kinds of the singularities, points and strings.  In the
"stringful" region, it is rather difficult to locate the
singularity \emph{point}.  It is because the operation of the
adiabatic process changes when making the contour across a string.
In Fig.~\ref{total_adiabatic}, we illustrate two different
adiabatic processes regarding making contours enclosing the
singularity points and across the singularity string.
Fig.~\ref{total_adiabatic}(a) shows the usual definition of the
adiabatic process.  In the presence of the gap, the adiabatic
process has to follow the lower energy state, which is indicated
by the square dots in Fig.~\ref{total_adiabatic}(a).
Fig.~\ref{total_adiabatic}(b) shows the adiabatic process of the
contour across the string.  In this case, because the wave
functions of the HOMO and the LUMO do not overlap, the adiabatic
process needs to keep ramping up to the higher energy state at the
first degenerate point and ramp down to the lower energy state at
a second degenerate point, which completes a close loop.  Note
that the energy of the HOMO is larger than that of the LUMO after
the first degenerate point and becomes smaller again after the
second degenerate point.  In other words, the system stays in the
same state, and thus the charge does not transfer at all after a
close contour. Therefore, even if we make a close loop which
encloses a whole singularity string, the charge does not transfer.
It should be noted that even though the system is at a higher
energy state, it takes $\sim e^{L/\xi}$ to relax, where $L$ is the
size of the system, and $\xi$ is the typical localization length
which is around 3 sites in our case.

In Fig.~\ref{total_adiabatic}(c), we demonstrate a real process
across a singularity string indicated by the loop {\bf C} in
Fig.~\ref{fig:distribution}.  Usually, we refer singularity string
to the degenerate manifold between the HOMO and LUMO.  In a real
process, we may have to consider the degeneracy between the HOMO and
the 2$^{\text{nd}}$ LUMO.  For example, after the degenerate point
between LUMO and HOMO, the descending order in energy is
2$^{\text{nd}}$ LUMO, HOMO, and the LUMO. There is the possibility,
for example the boundary $\beta$ in Fig.~\ref{total_adiabatic}(c),
that the energy of 2$^{\text{nd}}$ LUMO and that of HOMO can be the
same along the contour before meeting the next degenerate point
between HOMO and LUMO.  If it happens, the system still stays in the
original HOMO state by ramping further up to the higher energy
state.  After that, the descending order in energy becomes HOMO,
2$^{\text{nd}}$ LUMO, and LUMO.  Fig.~\ref{total_adiabatic}(c) shows
the complete process and the system has to go back to its original
state after completing a close loop.  As a result, HOMO, LUMO, and
other states can be viewed as the Riemann sheets. Singularity
strings are like the branch cuts which connect different Riemann
sheets.  Contours on the $(Q_1, Q_2)$ plane are very
similar to the one in the complex plane.

As the string just corresponds to the degenerate states which does
not penetrate through the whole sample, the only topological
object which contributes to the polarization is the isolated
singularity \emph{point}.  We found one at $(Q_1,
Q_2,\alpha)=(-0.3885, - 4.80415,0)$. If we calculate
(\ref{torus_surface}) around that point, for example loop {\bf
B} in Fig.~\ref{fig:distribution}, we obtain $\Delta P = -1.0$. We
identify the singularity point which carries $\Delta P = -1.0$
with the antimonopole in contract to the monopole which carries
$\Delta P = 1.0$.  Considering together with the total flux
calculated from the loop {\bf A} in Fig.~\ref{fig:distribution} to
be 1.0, \emph{there must be at least two monopoles and one
antimonopole in our system, which demonstrates the proliferation
of the monopole-antimonopole pairs in the strong disorder limit.}
Unfortunately, we are not able to locate monopoles exactly but
obtain rough position, since they merge in the string-ful soup.
The integral in (\ref{gauge_flux}) converges very slowly as the
counters are close to the singularity string (or points).  The
integral step is equivalent to the rate of the adiabatic process,
which has to be less than the typical energy spacing
$2(t+v_i^{\text{max}})/N$. When approaching to the singularity,
the energy gap becomes very small, so the process rate has to be
smaller to have convergent result. Therefore, it is very difficult
to locate the monopole position if they are in the string-ful
soup.

\begin{figure}[htbp]
  \begin{minipage}{4.2cm}
    \includegraphics[scale=0.2]{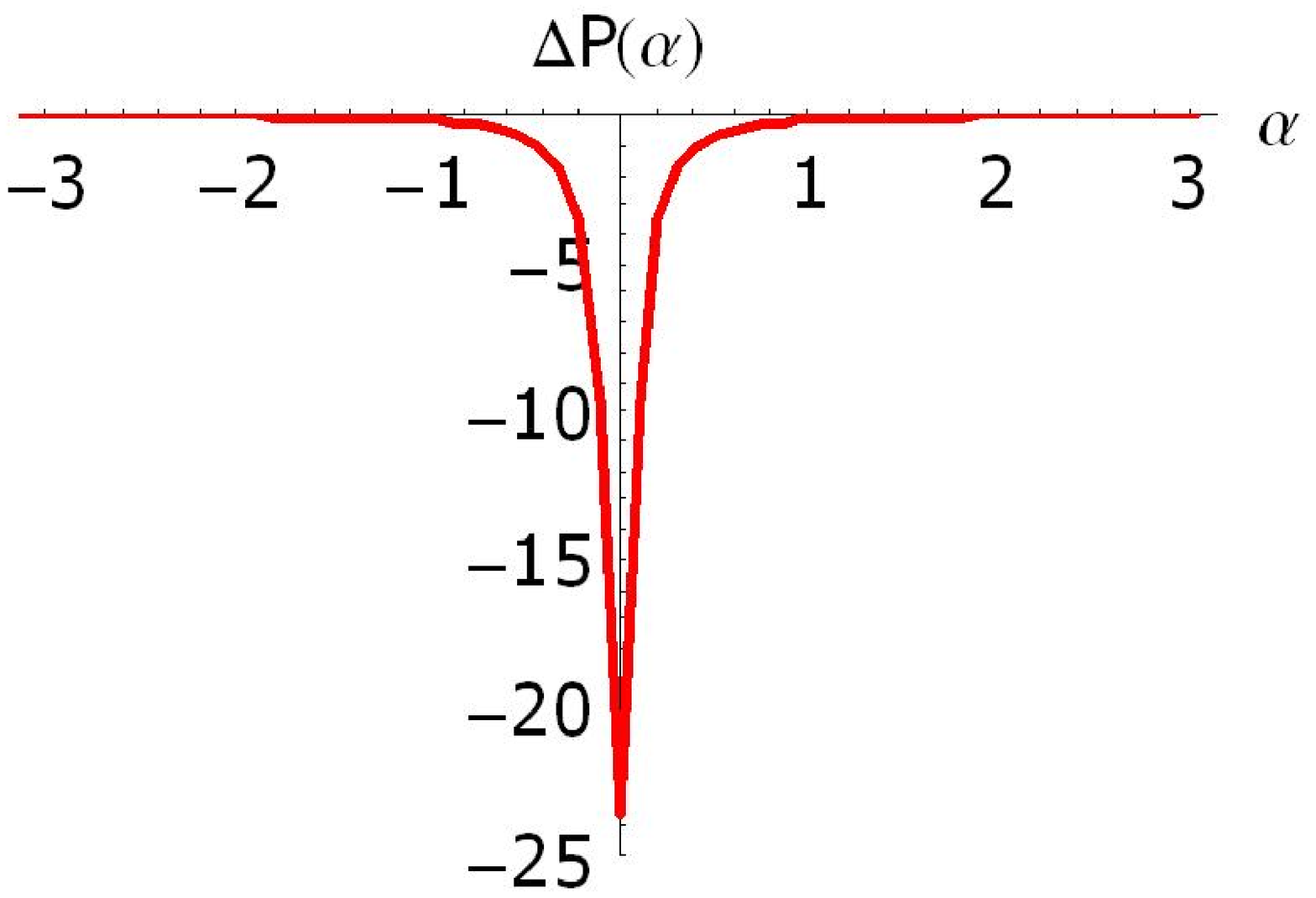}
    (a)
  \end{minipage}
  \begin{minipage}{4.2cm}
    \includegraphics[scale=0.2]{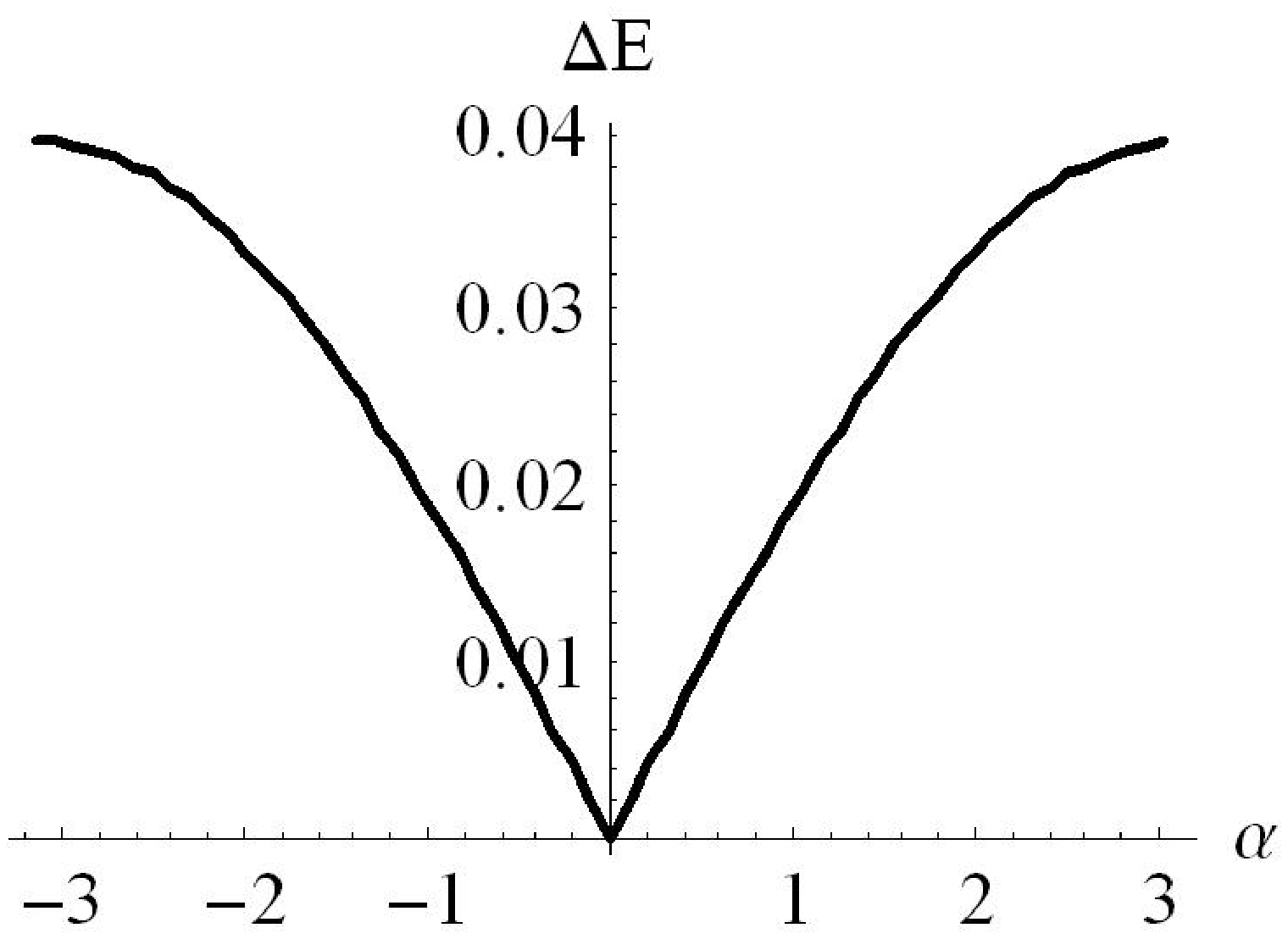}
    (b)
  \end{minipage}
  \begin{minipage}{4.2cm}
    \includegraphics[scale=0.2]{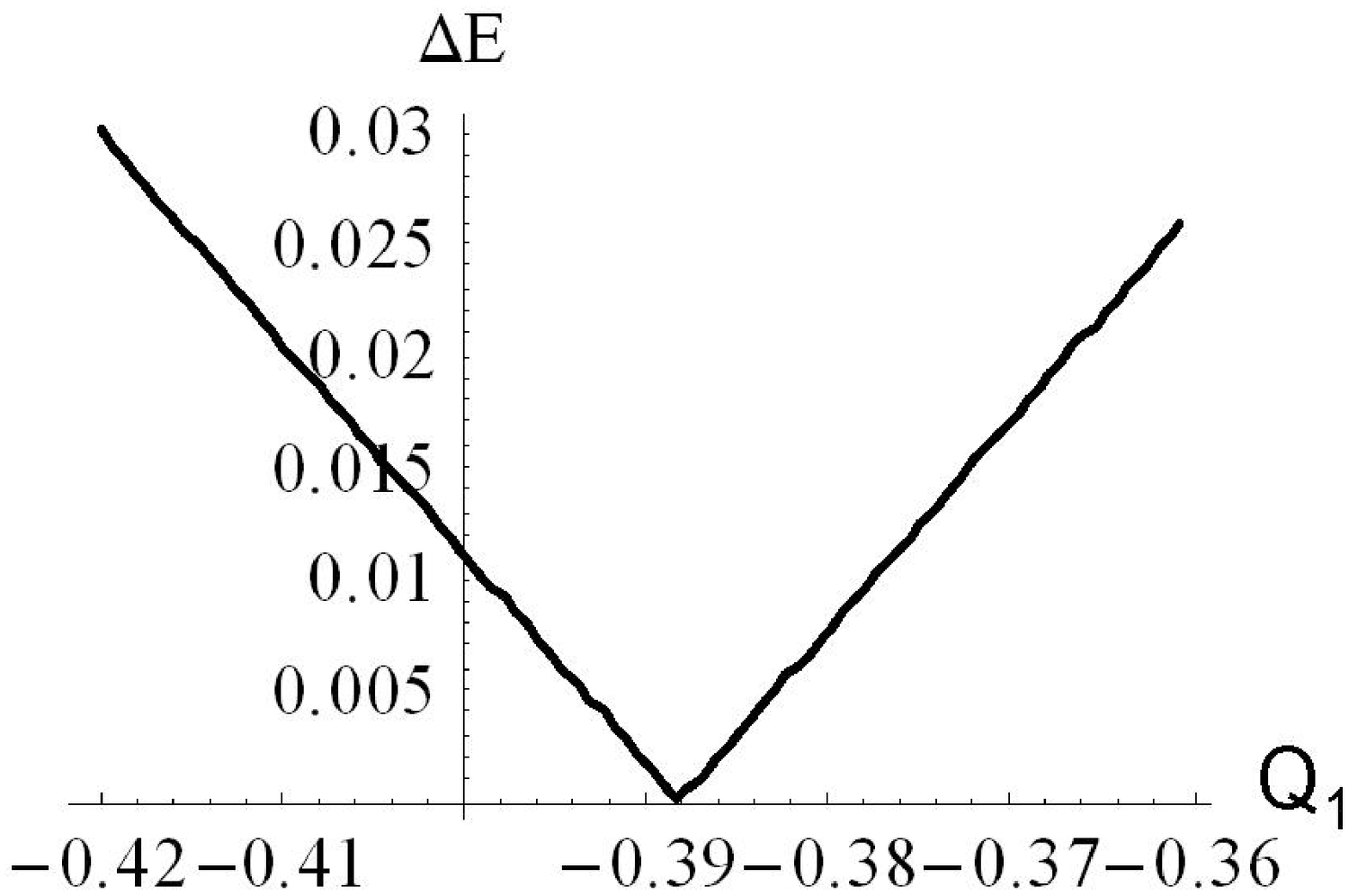}
    (c)
  \end{minipage}
  \begin{minipage}{4.2cm}
    \includegraphics[scale=0.2]{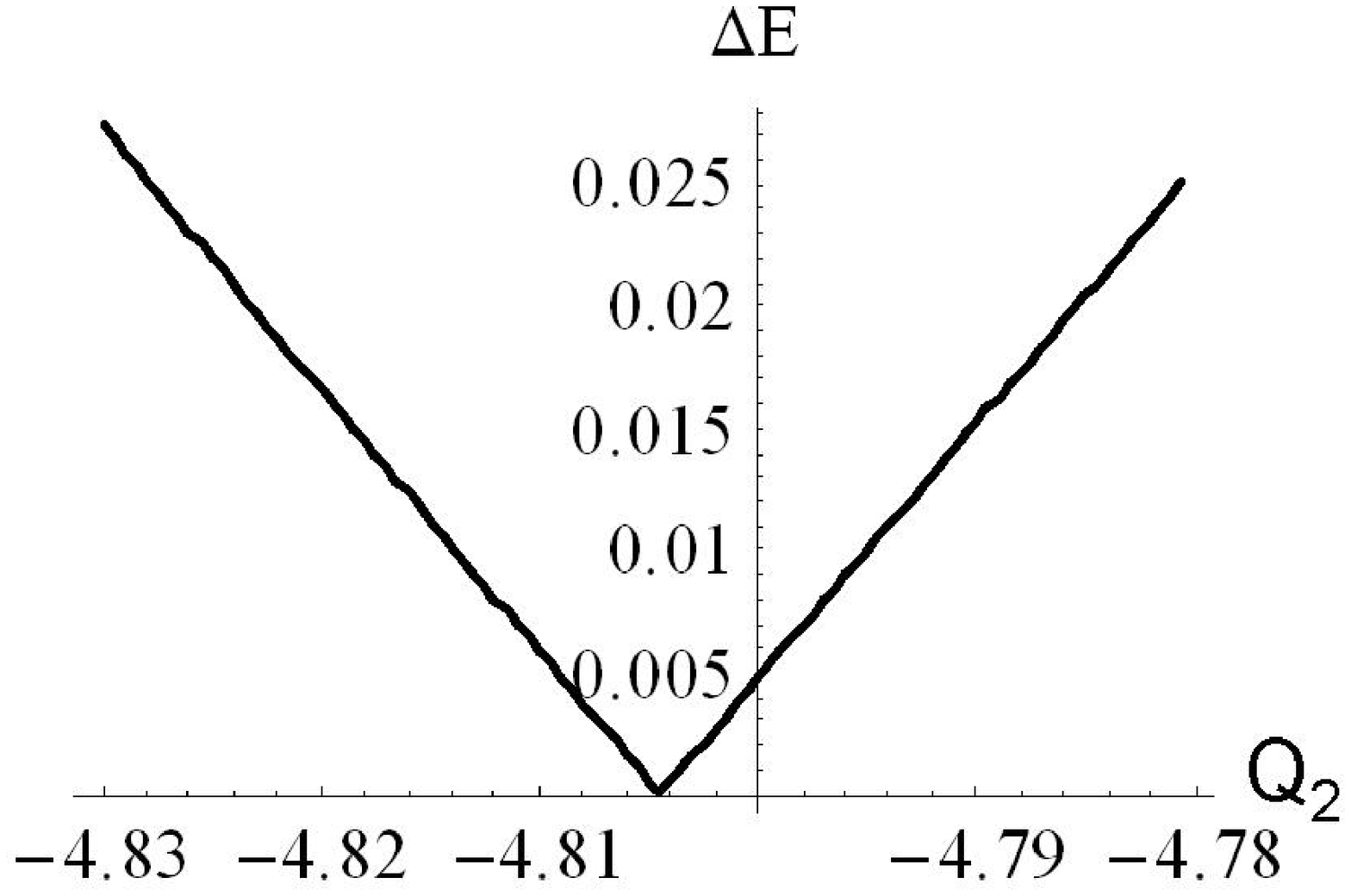}
    (d)
  \end{minipage}
  \begin{minipage}{4.2cm}
    \includegraphics[scale=0.2]{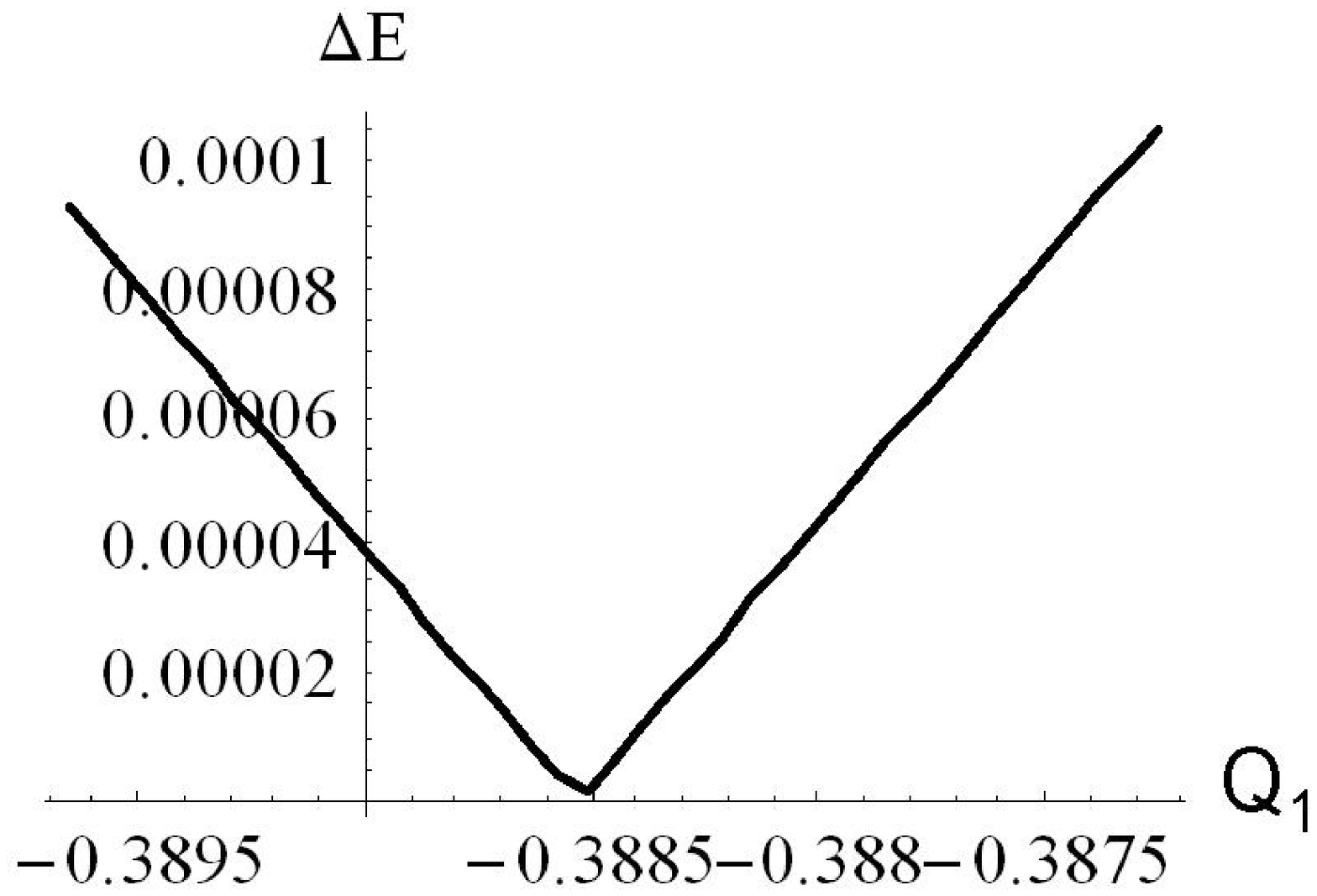}
    (e)
  \end{minipage}
    \begin{minipage}{4.2cm}
    \includegraphics[scale=0.2]{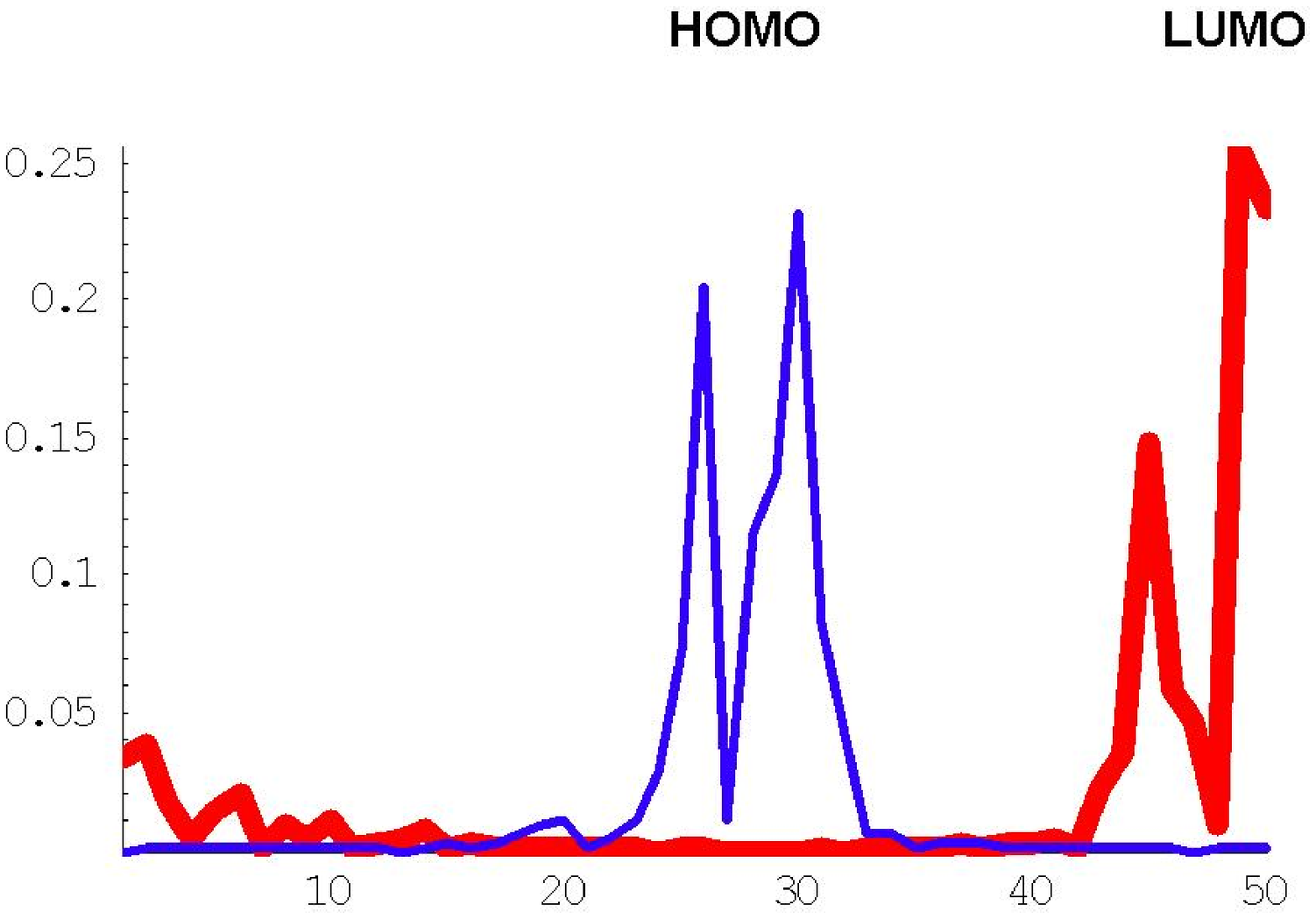}
    (f)
  \end{minipage}
\caption{\label{antimonopole} Some properties of antimonopole. (a)
the $\alpha$-dependence of (\ref{distribution}) around the
antimonopole.  (b), (c), and (d) the energy gap functions of the
antimonopole along the $\alpha$, $Q_1$, and $Q_2$ direction. (d)
the gap function along the valley.  The antimonopole is not
spherically symmetric.  The gap function at $\alpha=0$ has a deep
valley because the slopes along the valley is 10 times smaller
than other directions. (f) The wave functions of the HOMO and LUMO
at the antimonopole. In Figs.~(b)-(e) the axes are in the unit
of $t_{\text{n.n.}}$.}
\end{figure}

The appearance of the antimonopoles in the strong disorder limit
is very appealing.  It contributes to the polarization in the
opposite way to that a monopole does. The polarization comes from
the charge transportation through the whole sample from one end to
the other. Therefore, (\ref{distribution}) for the antimonopole
must be highly $\alpha$-dependent shown in the
Fig.~\ref{antimonopole}(a). The peak value of $\Delta P(\alpha)$
happens at $\alpha = 0$. After integrating over $\alpha$ by
(\ref{torus_surface}), we obtained $\Delta P = -1.0$.  This
result indicates that the antimonopole may be located at $\alpha =
0$.  In Fig.\ref{antimonopole}(b), we found that the energy gap
closes at $\alpha = 0$ supporting this observation. In
Fig.~\ref{antimonopole}(c) and \ref{antimonopole}(d), we show the
energy gap at $\alpha=0$ along $Q_1$ and $Q_2$ direction,
respectively.  In fact, the structure of the antimonopole is
highly anisotropic.  The energy gap has a valley-like structure at
$\alpha = 0$. Fig.~\ref{antimonopole}(e) shows the energy gap
along the valley. The slope of the energy gap function along the
valley is roughly 10 times smaller than other directions. In other
words, the antimonopole looks like a football more than a round
ball. This anisotropy can be understood in terms of the resonance
tunnelling as we see in the next section.

\begin{figure}[htbp]
  \includegraphics[scale=0.4]{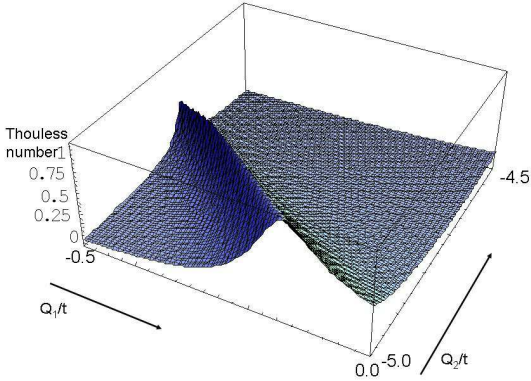}
  \caption{\label{fig:thouless_number}
The Thouless number in the vicinity of the antimonopole.
 The Thouless number has a ridge structure which exactly coincides with the valley
  in the gap function.}
\end{figure}

Another quantity to measure the extendedness of a state is the
Thouless number defined by
\begin{eqnarray}
\mathcal{N}_T = \frac{E(\alpha=\pi)-E(\alpha=0)}{\delta E}
\end{eqnarray}
where $E(\alpha)$ is the energy of the state and $\delta E$ is the
typical energy spacing.  We plot the Thouless number  of the HOMO
in Fig.~\ref{fig:thouless_number}. We found a ridge
distribution which coincides with the valley in the energy gap
function. The antimonopole locates at the ridge but not
necessarily the highest one.  The Thouless number at the ridge is
one order of magnitude larger than the other points in the
vicinity in the parameter space.  It suggests that the high
boundary condition sensitivity must dominate the physics at the
antimonopole.  It leads to the resonance tunnelling in the
Anderson insulator.\cite{Azbel1983}

\subsection{Resonance Tunnelling} \label{section:resonance-periodic}
We now compare the numerical results with the picture of resonance
tunnelling discussed in Sec.~\ref{section:resonance-open}. In the
resonance tunnelling scenario, we can deal with the periodic
system by identifying $x_3$ with $x_1$. It leads to $k_{3}=k_{1}$.
In such a periodic system, as is apart from an open system,
$E_{F}$ is no longer a tunable parameter, but corresponds to
discrete energy levels of the system. Thus, $S_{i}$, $\alpha_{i}$,
and $\beta_{i}$ are smooth functions of $Q_{1}$, $Q_{2}$ and an
energy $E$. Because the periodic system is threaded by a flux of
$\alpha{\Phi}_{0}$, we require that
\begin{equation}
\Theta
\left(\begin{array}{c}
A_{3}\\B_{3}\end{array}\right)
=\left(\begin{array}{c}
A_{1}\\B_{1}\end{array}\right)
=e^{i\alpha}\left(\begin{array}{c}
A_{3}\\B_{3}\end{array}\right).
\label{eq:periodiceq}\end{equation}
It yields ${\rm det}(\Theta-e^{i\alpha})=0$, i.e. $
{\rm Re}
\theta_{11}=\cos\alpha$. Because $\theta_{11}$ is
a function of $Q_{1}$, $Q_{2}$ and $E$, this defines a (discrete)
energy level of
the system as a function of $Q_1$ and $Q_2$.

\begin{figure}[htb]
\includegraphics[width=8.5cm]{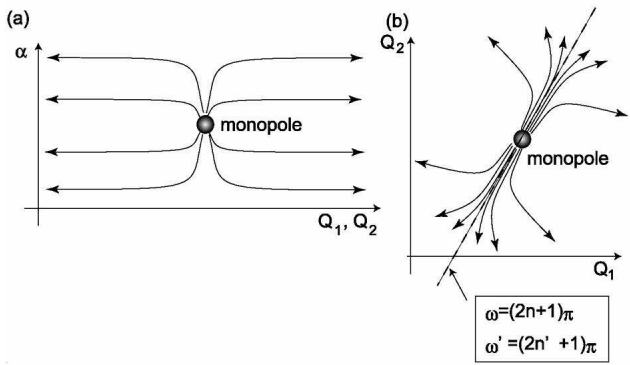}
\caption{Distribution of the vector field $\vec{F}$ around the monopole.
(a) the field $\vec{F}$ is mostly uniform in the $\alpha$-direction because
$\xi\ll L$, except for the vicinity of the monopole. (b) the field
is anisotropic in the $Q_1$-$Q_2$ plane.}
\label{fig:resonance-periodic}\end{figure}

We study the monopoles in the framework of resonance tunnelling.
As the monopoles appear at band-crossing points, we have to look
for degenerate solutions of (\ref{eq:periodiceq}). Namely, we
look for cases with two (degenerate) states for fixed $Q_1$, $Q_2$
and $E$. This happens if and only if the matrix $\Theta$ is equal
to $e^{i\alpha}$. This condition reduces to
\begin{eqnarray}
&&S_{1}=S_{2}, \label{eq:S1S2periodic}\\
&&\omega\cong 2\int_{x''_{1}}^{x'_{2}}k(x)dx =(2n+1)\pi, \label{eq:omegaperiodic}\\
&&\omega' \cong 2\int_{x''_{2}}^{x_{3}}k(x)dx +
2\int_{x_{1}}^{x'_{1}}k(x)dx
=(2n'+1)\pi, \label{eq:omega'periodic}\nonumber \\ \\
&&\alpha=-(n+n'+1)\pi, \label{eq:alphaperiodic}
\end{eqnarray}
where $n$ and $n'$ are integers. Because $S_1$, $S_2$, $\omega$ and
$\omega'$ are functions of $Q_1$, $Q_2$ and $E$, the three
conditions (\ref{eq:S1S2periodic}), (\ref{eq:omegaperiodic}) and
(\ref{eq:omega'periodic}) determine a set of isolated points in the
$Q_1$-$Q_2$-$E$ space. Combining with
(\ref{eq:alphaperiodic}), we get isolated points in the
$Q_1$-$Q_2$-$\alpha$ space, corresponding to the
monopoles/antimonopoles \cite{Cohen2003}. Except for the vicinity of the
monopoles/antimonopoles, the field $\vec{F}$ is almost $\alpha$-independent,
because the system is almost intact with the change of $\alpha$.
Thus the distribution of the field $\vec{F}$ is as shown
in Fig.~\ref{fig:resonance-periodic}(a).
In contrast with a small-sized
model $\xi\sim L$ in ref.~\onlinecite{Cohen2003},
in our disordered model ($\xi\ll L$)
the ``near-field'' region where the flux density is $\alpha$-dependent is
limited only in the very
close vicinity of the monopoles/antimonopoles.

Equations (\ref{eq:omegaperiodic}) and (\ref{eq:omega'periodic})
are the Bohr quantization conditions, meaning that the there are
two localized states around $x_{2}$ and $x_{3}$ ($\equiv x_{1}$),
having the same energy. In general, when there are two localized
states with the same energy, displaced each other by a high
potential peak, there is a small tunnelling matrix element between
them. It gives a small energy splitting between bonding and
anti-bonding states. As a result, the degeneracy is lifted.
Nevertheless, when the equality (\ref{eq:S1S2periodic}) holds, it
guarantees that the hybridization between them cancels exactly,
and the degeneracy is not lifted even after one takes the
tunnelling into account. Thus the two localized states are in
resonance with each other. Equation (\ref{eq:alphaperiodic}) means
that such exact degeneracy occurs only when $\alpha$ is equal to
$\pi$ or $0$, confirming our numerical result.

If we change $Q_{1}$ and $Q_{2}$ to make  $S_{1}\neq S_{2}$ with
the conditions (\ref{eq:omegaperiodic}) and (\ref{eq:omega'periodic})
preserved, there occurs a small splitting ($\sim O(e^{-2S_{i}})$)
to the otherwise degenerate
states, due to a small unbalance between $S_1$ and $S_2$.
These conditions (\ref{eq:omegaperiodic}) and (\ref{eq:omega'periodic})
specify one direction in the $Q_1$-$Q_2$ plane from the monopole.
Along this direction the gap remains nonzero but very small, thereby
the field $\vec{F}$ becomes large as schematically shown
in Fig.~\ref{fig:resonance-periodic} (b).
In our numerical results,
this direction corresponds to
the valley direction in the energy gap, and the ridge in
the Thouless number in Fig.~\ref{fig:thouless_number}.
Thus the anisotropy of the energy gap in the $Q_1$-$Q_2$ plane is of the
order $e^{2S_{i}}\sim e^{L/\xi}$.
Because
in our numerical calculation $L=50$, $\xi\sim 7$, the anisotropy of the
monopole is $e^{2S_{i}}\sim e^{L/\xi}\sim e^{7}\sim 10^{3}$,  which
should be compared with the numerical value $\sim 10$ in the previous
subsection. Their difference can be attributed to finiteness of the mesh
size in the numerical calculation; the mesh may not be fine enough
to reproduce the anisotropy $\sim 10^3$.
We also note that the strings observed in the numerical calculation
can be regarded as the curve of $\omega=(2n+1)\pi$, $\omega'=(2n'+1)\pi$,
with $S_1$ and $S_2$ are unrestricted.

\subsection{Discussion}
In Fig.(\ref{antimonopole}f), we show the wave functions of the
HOMO and LUMO, which supports the resonance tunnelling scheme. At
the resonance, the energy of HOMO and LUMO are degenerate.  If
HOMO sits at $x_1$, LUMO can be viewed as the resonance state at
the potential valley $x_2$.  Therefore, by means of LUMO, HOMO can
tunnel through the whole sample.  We must emphasize that this
mechanism is topological as that in the pure case.

The adiabatic rate depends on the energy gap.  At the resonance, the
wave functions of the HOMO and the LUMO do not overlap.  By tuning
the parameters, the overlapping integral between these two states
are not zero, and thus the gap opens up.  Again, the rates that the
gap opens with the parameters are not necessarily the same.
Therefore, the monopole (antimonopole) structure is not necessarily
isotropic.  If fact, it can be highly anisotropic as shown in our
calculation.  Then, the adiabatic rate is restricted by the minimal
gap along the contour.  It can be roughly estimated as
$e^{-L/\xi_R}$, where $\xi_R$ is the localization length at the
resonance.

The current results for the periodic/twisted boundary condition is
consistent with those for the open boundary condition. The
topological structure in the open boundary condition, namely the
vortices, can be regarded as a projection of the monopoles to a
2-dimensional plane.  The anisotropic flux distribution and the
anisotropic phase gradient are two facets of the resonance
tunnelling. While the results for the open boundary condition is
more applicable to the real materials and design, the non-trivial
generation of the monopole-anti-monopole shown in the
periodic/twisted boundary condition case illustrates the deep and
whole new topological structure in the strong disordered systems.

\section{Conclusion}

In this paper, we have studied the charge pumping and dielectric
response in disordered insulator for both the open boundary
condition and periodic/twisted boundary condition. As for the open
system coupled to the leads, we have found the
quantum-mechanically enhanced dielectric response in
nanoscopic/mesoscopic disordered insulators, which give a guide
for fast and low-dissipation ferroelectric thin films of the
FeRAM. The phase of the reflection coefficient $r$ is a key
parameter for the pumping, and has the rich structure in the
parameter plane. Topological nature of the insulator dictates the
phase winding of $r$, around the vortex where $r=0$ and $|t|=1$.
In a pure insulator, this corresponds to the gapless point. In the
disordered case, it corresponds to the resonance tunnelling
through the sample, whose position varies with the chemical
potential.

With the periodic closed system, we can consider the three
dimensional parameter space ${\vec Q} = (Q_1, Q_2, \alpha)$ where
$Q_1$ is the site energy alternation, $Q_2$ is the bond
dimerization, and $\alpha$ is the phase twist for the boundary
condition. In this three dimensional space, one can discuss the
monopoles and the their associated gauge field distribution, which
are again interpreted in terms of the resonant tunnelling.

The existence of the resonant tunnelling in the parameter space is
guaranteed by the topological properties of the charge pumping,
and only two parameters such as $Q_1$ and $Q_2$ need to be tuned
to realize it. These parameters can be controlled experimentally
by the external electric field and pressure. Therefore the present
theory offers a way to design the enhanced dielectric response in
realistic systems by controlling the disorder.

\begin{acknowledgments}
The authors would like to thank D. Vanderbilt, S. Horiuchi, Y.
Okimoto, Y. Ogimoto, and Y. Tokura for stimulating discussion. The work was
partly supported by Grant-in-Aids under the Grant numbers
15104006, 16076205, and 17105002, and NAREGI Nanoscience Project
from the Ministry of Education, Culture, Sports, Science, and
Technology.
\end{acknowledgments}

%\bibliography{cdw}

\end{document}